\documentclass[aps,prl,twocolumn,citeautoscript,superscriptaddress]{revtex4-2}
\pdfoutput=1

\usepackage{natbib}
\usepackage[english]{babel}
\usepackage{letltxmacro}
\usepackage{latexsym}
\LetLtxMacro{\ORIGselectlanguage}{\selectlanguage}
\makeatletter
\DeclareRobustCommand{\selectlanguage}[1]{%
  \@ifundefined{alias@\string#1}
    {\ORIGselectlanguage{#1}}
    {\begingroup\edef\x{\endgroup
       \noexpand\ORIGselectlanguage{\@nameuse{alias@#1}}}\x}%
}
\newcommand{\definelanguagealias}[2]{%
  \@namedef{alias@#1}{#2}%
}
\makeatother
\definelanguagealias{en}{english}
\definelanguagealias{English}{english}
\usepackage{graphicx}
\usepackage{amsmath}
\usepackage{amsfonts}
\usepackage{amssymb}
\usepackage{bm}
\usepackage{color}
\usepackage[percent]{overpic}
\usepackage{soul} 
\usepackage{amssymb}
\usepackage{wasysym}
\usepackage{dsfont}
\usepackage{float}
\usepackage[mathscr]{euscript}
\usepackage{lipsum}
\usepackage{hyperref}
\hypersetup{
    pdfstartview={FitH},    % fits the width of the page to the window
    colorlinks=true,       % false: boxed links; true: colored links
    linkcolor=blue,          % color of internal links (change box color with linkbordercolor)
    citecolor=blue,        % color of links to bibliography
    filecolor=magenta,      % color of file links
    urlcolor=blue           % color of external links
}

\newcommand{\pdagger}{\phantom{\dagger}}
\newcommand{\be}{\begin{equation}}
\newcommand{\ee}{\end{equation}}
\newcommand{\bea}{\begin{eqnarray}}
\newcommand{\eea}{\end{eqnarray}}

\newcommand{\ket}[1]{| {#1} \rangle}

\usepackage{graphicx}
\usepackage[colorinlistoftodos]{todonotes}
\usepackage{verbatim}
\usepackage[normalem]{ulem}

\newcommand{\Tr}{\mathrm{Tr}}
\usepackage{pdfpages}
\usepackage{pgffor}
\makeatletter
\AtBeginDocument{\let\LS@rot\@undefined}
\makeatother

\begin{document}

\title{Quantum turnstiles for robust measurement of full counting statistics}

\author{Rhine Samajdar}

\affiliation{Department of Physics, Princeton University, Princeton, NJ 08544 }
\affiliation{Princeton Center for Theoretical Science, Princeton University, Princeton, NJ 08544}

\author{Ewan McCulloch}

\affiliation{Department of Physics, University of Massachusetts, Amherst, MA 01003}

\author{Vedika Khemani}

\affiliation{Department of Physics, Stanford University, Stanford, CA 94305}

\author{Romain Vasseur}

\affiliation{Department of Physics, University of Massachusetts, Amherst, MA 01003}

\author{Sarang Gopalakrishnan}

\affiliation{Department of Electrical and Computer Engineering, Princeton University, Princeton, NJ 08544}

\begin{abstract}
We present a scalable protocol for measuring full counting statistics (FCS) in experiments or tensor-network simulations. In this method, an ancilla in the middle of the system acts as a turnstile, with its phase keeping track of the time-integrated particle flux. Unlike quantum gas microscopy, the turnstile protocol faithfully captures FCS starting from number-indefinite initial states or in the presence of noisy dynamics. In addition, by mapping the FCS onto a single-body observable, it allows for stable numerical calculations of  FCS using approximate tensor-network methods. We demonstrate the wide-ranging utility of this approach by computing the FCS of the transferred magnetization in a Floquet Heisenberg spin chain, as studied in a recent experiment with superconducting qubits, as well as the FCS of charge transfer in random circuits.
\end{abstract}
\maketitle

\textit{Introduction}.---Conventional linear-response transport probes how the expectation values of densities and currents respond to external perturbations. 
In general, the point of transport experiments is to shed light on the intrinsic dynamics of a system. However, in many cases, theoretically natural questions about (say) the charge or the lifetime of a system's elementary excitations cannot be resolved at the level of linear response. 
To settle these questions, one must probe higher-order correlation functions or depart from equilibrium, using techniques such as nonlinear response~\cite{PhysRevLett.122.257401, PhysRevB.99.045121, PhysRevB.102.165137, mahmood2021observation, fava2021hydrodynamic, PhysRevResearch.3.013254, PhysRevB.103.L201120, PhysRevLett.128.187402, PhysRevLett.128.076801, mcginley2022signatures} or noise spectroscopy~\cite{blanter2000shot, kolkowitz2015probing, sinitsyn2016theory, rovny2022nanoscale}. 

An especially informative way of capturing these higher-order correlations is full counting statistics (FCS), i.e., the full distribution function of charge transferred across a cut. FCS was initially introduced in the context of mesoscopic systems \cite{Levitov1993,Levitov1996Electron,Belzig2001,Belzig2002,Levitov_2004,Gogolin_2006,Levitov2003}, but contemporary experimental platforms built on synthetic matter, such as ultracold atomic gases or superconducting qubit arrays, allow for much more direct access to this observable. For instance, one can take simultaneous snapshots of all the particles in a quantum gas microscope experiment \cite{Bakr_2009,Sherson2010,bloch2012,Haller2015,Islam2015,Parsons_2016,Choi2016,Mazurenko2017, Gritsev_2006, Hofferberth_2008}, which, in turn, yield the FCS. While early studies of FCS were focused on low-temperature properties, recent work has explored how the FCS evolves after quenches \cite{Ashida_2018,Ridley2018,Ridley2019,Bastianello_2018,McCulloch2023,Bertini2023,XXZFCS,Wei2022}; a notable finding is that in integrable models as well as certain constrained nonintegrable models, the FCS remains strongly non-Gaussian at all times \cite{Wei2022, XXZFCS, DeNardis2022,Krajnik2022a,Krajnik2022b,Krajnik2022c,PhysRevLett.123.200601,PhysRevLett.128.090605}. Many basic questions remain unanswered, however, including the conditions under which the FCS of quantum fluctuations has an effective classical description.

\begin{figure}[t]
    \centering
    \includegraphics[width=\linewidth,trim= 0 6cm 0 0, clip]{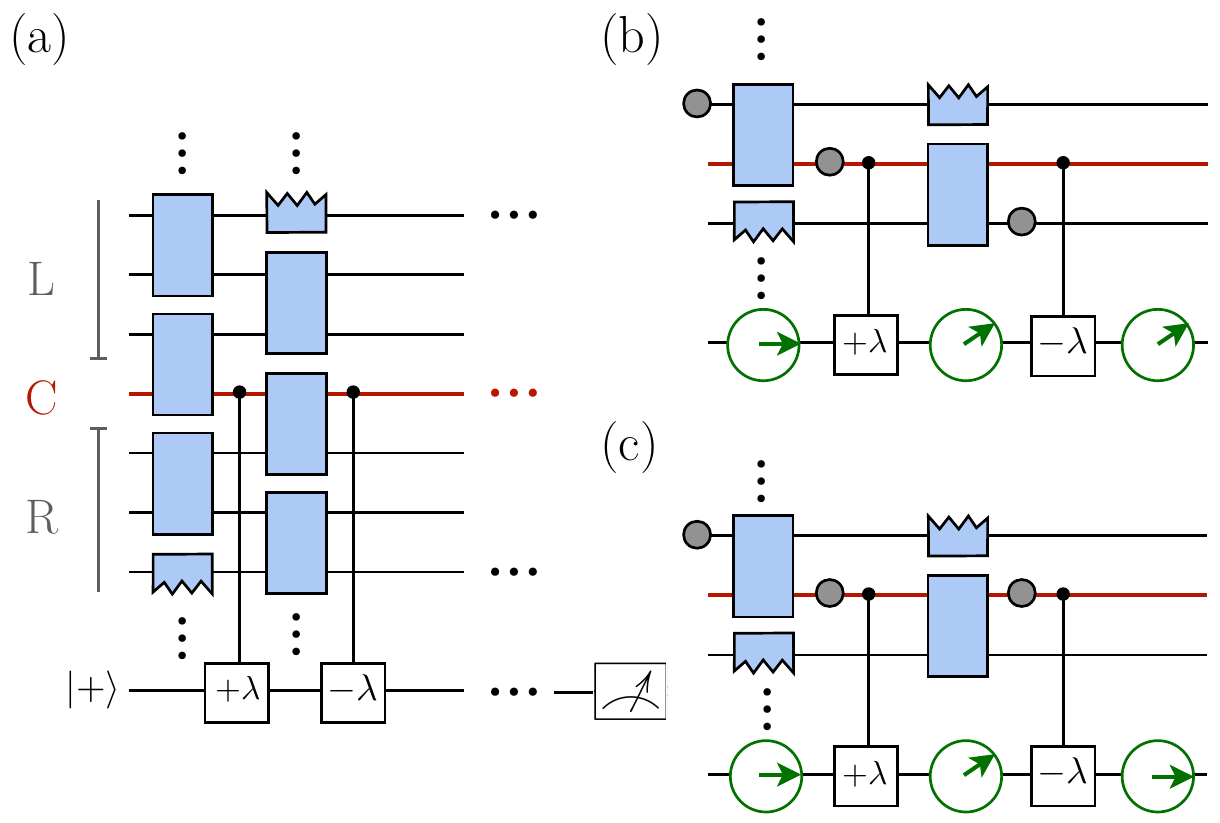}
    \caption{(a) Implementation of the turnstile for one cycle of the brickwork circuit. The controlled phase gates acting on the ancilla are denoted by their respective angles $\pm \lambda$. (b, c) As a particle (gray circle) hops on and off the central site, labeled ``C'', the change in its occupation is imprinted on the ancilla (as schematically depicted by the green dial) and can be related to the current as discussed below.}
    \label{fig:idea}
\end{figure}

So far, experiments probing FCS in synthetic matter have been carried out by initializing a system in a state with a sharp number of particles in the subsystem of interest \cite{Wei2022}, letting the system evolve, and then extracting the particle-number distribution from many snapshots. Despite its straightforwardness, this technique has intrinsic limitations. First, it is limited to initial states with a sharp number of particles in the subsystem of interest, a restriction that eliminates most thermal equilibrium states. Second, in the presence of noise, its performance scales exponentially poorly with subsystem size, since the particle number count is vulnerable to decoherence deep inside the subsystem. Third, the FCS involves arbitrary-order correlation functions, which are difficult to retain faithfully in approximate numerical algorithms based on tensor networks. 

In the present work, we propose and explore a method that circumvents these obstacles. This method is essentially a concrete implementation of the thought experiment in Ref.~\onlinecite{Levitov1996Electron} but exploits a further simplification due to the brickwork circuit structure. It involves an ancilla that acts as a turnstile: the net flux of particles across the turnstile is imprinted on the phase of the ancilla, and the FCS can be read off from this phase information. By tracking the flux instead of the full particle number, this approach evidently overcomes the drawbacks of the direct-counting approach: it is robust against decoherence deep inside the subsystem, and also does not rely on using number-definite initial states. While interferometry using ancillas is a familiar technique for measuring nontrivial many-body observables~\cite{PhysRevA.86.032324}, including FCS \cite{xu2019probing}, our proposal is distinctive in using only \textit{local} two-site gates, and it can therefore be realized with minimal overhead in near-term experiments. This feature also makes it practical to implement in approximate numerical algorithms: mapping the FCS onto a single-site expectation value opens the door to its computation using tensor-network methods, such as the density matrix truncation algorithm \cite{White2018}, which are designed to preserve few-site expectation values but not high-order correlations.

\textit{Protocol.}---We consider the brickwork unitary circuit geometry illustrated in Fig.~\ref{fig:idea}(a) acting on a chain of $N$ qubits with conserved particle number (interpreting the states $\ket{0}$ and $\ket{1}$ as empty and occupied, respectively). We are interested in particle transport and fluctuations between two halves of the system, starting from an arbitrary (possibly non-number-sharp) initial state.
One can regard this, alternatively, as a Trotterized nearest-neighbor Hamiltonian but the unitary gates can have arbitrary time dependence. Each layer of gates is bracketed by two ``turnstile'' gates involving the central qubit (labeled in red) and an ancilla, which is initialized in the state $\ket{+} \equiv 2^{-1/2}(\ket{0} + \ket{1})$. The turnstile gates apply rotations by an angle $\pm \lambda$ about the $z$ axis on the ancilla, controlled on the central qubit being in the state $\ket{1}$. The intuition behind the protocol is as follows. The net effect of the turnstile gates before and after the layers depicted in Fig.~\ref{fig:idea}(a) is to imprint a phase $\lambda (-\lambda)$ if the occupation of the central site went down (up) during that cycle. Crucially, because of the brickwork geometry, any particle that hops on (off) the central site during this layer must have come from (or gone to) sites \emph{above} it. During the next layer, likewise, any change in the occupation of the central site corresponds to transport between the central site and those \emph{below} it. To measure the current, it suffices for the ancilla to keep track of the change in the occupation of the central site.

We now lay this idea out more formally. Suppose that after a certain number of cycles, the system is in the state
\begin{equation*}
\rvert \psi \rangle =  \sum_{s^{}_{\textsc{l}}, s^{}_{\textsc{c}}, s^{}_{\textsc{r}}} c^{}_{s^{}_{\textsc{l}}, s^{}_{\textsc{c}}, s^{}_{\textsc{r}}} \rvert s^{}_{\textsc{l}}, s^{}_{\textsc{c}}, s^{}_{\textsc{r}} \rangle^{}_{\mathrm{data}} \otimes \rvert \varphi (s^{}_{\textsc{l}}, s^{}_{\textsc{c}}, s^{}_{\textsc{r}}) \rangle^{}_{\mathrm{ancilla}},
\end{equation*}
where the sum runs over all configurations of ``data'' qubits on the left, center, and right parts of the chain, denoted by $s^{}_{\textsc{l}}$, $s^{}_{\textsc{c}}$, and $s^{}_{\textsc{r}}$, respectively, and $\rvert \varphi \rangle^{}_{\mathrm{ancilla}} \equiv$ \,$(\rvert 0 \rangle_{\mathrm{ancilla}} + e^{i \varphi} \rvert 1\rangle_{\mathrm{ancilla}})/\sqrt{2}$. By virtue of linearity, it is sufficient to examine one particular term in the superposition,
$
\rvert \psi^{}_0 \rangle$\,$=$\,$\left\rvert \bar{s}^{}_{\textsc{l}}, \bar{s}^{}_{\textsc{c}},\bar{s}^{}_{\textsc{r}} \right\rangle^{}_{\mathrm{data}} \otimes \rvert \varphi \rangle^{}_{\mathrm{ancilla}},
$
where $\bar{s}^{}_{\textsc{l}}$\,$\left(\bar{s}^{}_{\textsc{r}}\right)$ represents a specific configuration that has a total of, say, $n_{\textsc{l}}$\,($n_{\textsc{r}}$) particles. Let us further assume, without loss of generality, that  $\bar{s}^{}_{\textsc{c}}$\,$=$\,$\rvert 0 \rangle$. After half a cycle, i.e., one layer of unitaries and controlled phase gates, the state $\rvert \psi^{}_0 \rangle$ evolves to
\begin{alignat}{1}
\rvert \psi^{}_1 \rangle &= \sum_{s^{(0)}_{\textsc{l}}, s^{(0)}_{\textsc{r}}}  \alpha^{}_{s^{(0)}_{\textsc{l}}, s^{(0)}_{\textsc{r}}} \left\rvert s^{(0)}_{\textsc{l}}, 0, s^{(0)}_{\textsc{r}} \right\rangle^{}_{\mathrm{data}}  \otimes \rvert \varphi \rangle^{}_{\mathrm{ancilla}}
\\&+\sum_{s^{(-1)}_{\textsc{l}}, s^{(0)}_{\textsc{r}}}  \alpha^{}_{s^{(-1)}_{\textsc{l}}, s^{(0)}_{\textsc{r}}} \left\rvert s^{(-1)}_{\textsc{l}}, 1, s^{(0)}_{\textsc{r}} \right\rangle^{}_{\mathrm{data}}  \otimes \rvert \varphi + \lambda \rangle^{}_{\mathrm{ancilla}} \nonumber,
\end{alignat}
where $s^{(m)}_{\textsc{l},\textsc{r}}$ stands for all configurations of the left/right data qubits with $n^{}_{\textsc{l},\textsc{r}}$\,$+$\,$m$ total particles. Subsequently, after one full cycle, the state of the system is given by
\begin{widetext}
\begin{alignat}{4}
\nonumber
\rvert \psi^{}_2 \rangle &= \sum_{s^{(0)}_{\textsc{l}}, s^{(0)}_{\textsc{r}}}  \alpha^{}_{s^{(0)}_{\textsc{l}}, s^{(0)}_{\textsc{r}}} \left\rvert s^{(0)}_{\textsc{l}}, 0, s^{(0)}_{\textsc{r}} \right\rangle^{}_{\mathrm{data}}  \otimes \rvert \varphi \rangle^{}_{\mathrm{ancilla}}
+\sum_{s^{(0)}_{\textsc{l}}, s^{(-1)}_{\textsc{r}}}  \alpha^{}_{s^{(0)}_{\textsc{l}}, s^{(-1)}_{\textsc{r}}} \left\rvert s^{(0)}_{\textsc{l}}, 1, s^{(-1)}_{\textsc{r}} \right\rangle^{}_{\mathrm{data}}  \otimes \rvert \varphi - \lambda \rangle^{}_{\mathrm{ancilla}}\\
&+\sum_{s^{(-1)}_{\textsc{l}}, s^{(0)}_{\textsc{r}}}  \alpha^{}_{s^{(-1)}_{\textsc{l}}, s^{(0)}_{\textsc{r}}} \left\rvert s^{(-1)}_{\textsc{l}}, 1, s^{(0)}_{\textsc{r}} \right\rangle^{}_{\mathrm{data}}  \otimes \rvert \varphi \rangle^{}_{\mathrm{ancilla}}
+\sum_{s^{(-1)}_{\textsc{l}}, s^{(+1)}_{\textsc{r}}}  \alpha^{}_{s^{(-1)}_{\textsc{l}}, s^{(+1)}_{\textsc{r}}} \left\rvert s^{(-1)}_{\textsc{l}}, 0, s^{(+1)}_{\textsc{r}} \right\rangle^{}_{\mathrm{data}}  \otimes \rvert \varphi + \lambda \rangle^{}_{\mathrm{ancilla}}.
\label{eq:phase}
\end{alignat}
\end{widetext}
Clearly, the phase accrued by the ancilla qubit in Eq.~\eqref{eq:phase} counts the number of particles that have passed on to the right side of the chain. It is straightforward to see that a similar result holds  if the central qubit is initially in the state $\bar{s}^{}_{\textsc{c}}$\,$=$\,$\rvert 1 \rangle$. Hence, if the ancilla is initialized in the $\rvert + \rangle$ state, upon tracing out the data qubits, we find that 
$
\rho^{}_{\mathrm{ancilla}} = \sum_Q p^{}_Q \rvert Q \lambda \rangle \langle Q \lambda \lvert, 
$
where $Q$ is the number of transferred particles. Consequently, by measuring the \textit{single-site} observables
$
\langle X \rangle = \sum_Q p^{}_Q \cos(Q \lambda)$,  $\langle Y \rangle$\,$=$\,$\sum_Q p^{}_Q \sin(Q \lambda)
$
we can construct the generating function 
$
f(\lambda)$\,$=$\,$ \sum_Q p^{}_Q \exp(i Q \lambda), 
$
whereupon taking derivatives with respect to $\lambda$ in the limit $\lambda \rightarrow 0$ yields all the moments $\langle Q^m \rangle$ for integer $m$.

\textit{Closed-system numerics.}---Having established the theoretical framework behind our ancilla-assisted measurement of FCS, we now showcase this method in the context of a particular model, namely, the one-dimensional XXZ chain described by the Hamiltonian 
\begin{equation}
\label{eq:HXXZ}
H = - J \sum_{j = 1}^{N-1} \left(S^{x}_{j} S^{x}_{j+1} + S^{y}_{j} S^{y}_{j+1}+ \Delta S^{z}_{j} S^{z}_{j+1} \right),
\end{equation}
where $S^{\nu}$\,$=$\,$\sigma^\nu/2$ are spin-$1/2$ operators. We will focus on spin transport in the isotropic case of $\Delta$\,$=$\,$1$, whereupon Eq.~\eqref{eq:HXXZ} reduces to the Heisenberg (XXX) model.  While the XXZ model is known to be integrable, remarkably, it was recently shown that a particular Trotterized version of this model preserves integrability \cite{vanicat2018integrable,ljubotina2019ballistic}, which allows us to faithfully study XXZ dynamics via discrete time evolution.
In this formulation, the discrete-time Floquet dynamics are governed by the two-step propagator
\begin{equation*}
\label{eq:U}
\mathcal{U} = \mathcal{U}^{}_{\mathrm{even}} \mathcal{U}^{}_{\mathrm{odd}} = \prod_{j=1}^{N/2} U^{}_{2j-1,2j}(\theta, \phi)   \prod_{k=1}^{N/2} U^{}_{2k,2k+1} (\theta, \phi), 
\end{equation*}
where the half-step $\mathcal{U}^{}_{\mathrm{even}}$ ($ \mathcal{U}^{}_{\mathrm{odd}}$) updates all odd--even (even--odd) numbered pairs of spins with the unitary
\begin{equation}
U (\theta, \phi) = \begin{pmatrix}
1 & 0 & 0 & 0\\
0 & \cos \theta & -i \sin \theta & 0\\
0 &  -i \sin \theta & \cos \theta &0\\
0 & 0 & 0 & e^{i \phi}
\end{pmatrix}.
\end{equation}
We choose $\theta$\,$=$\,$0.4 \pi$ and $\phi$\,$=$\,$0.8 \pi$, which accordingly sets $\Delta$\,$=$\,$  \sin(\phi/2)/\sin(\theta)$\,$=$\,$1$. Adding in the ancilla qubit, we sandwich controlled phase gates with alternating angles $\pm \lambda$ in between the $\mathcal{U}^{}_{\mathrm{even}}$ and  $\mathcal{U}^{}_{\mathrm{odd}}$ layers.

As mentioned earlier, one of the central advantages of our method is that allows us to study spin transport and FCS starting with initial states that are not necessarily number sharp. Specifically, we consider a family of states with a domain wall between the left and right halves of the system, which are described by the density matrix 
\begin{equation}
\label{eq:rho}
\rho(t=0) \propto (e^{\mu S^z})^{\otimes N/2}\,  \otimes \,( e ^{-\mu S^z}) ^{\otimes N/2},
\end{equation}
with the height of the domain wall given by $\tanh \mu$.
As $\mu$\,$\rightarrow$\,$\infty$, all the spins in the left (right) half point up (down), and the system is furthest from equilibrium. On the other hand, at $\mu=0$, the initial state is the equilibrium thermal state at infinite temperature, and the SU(2) symmetry of the Heisenberg model is restored. This limit is particularly interesting from the viewpoint of quantum transport as the corresponding dynamics have been conjectured to be in the celebrated Kardar-Parisi-Zhang (KPZ) universality class \cite{znidarivc2011spin,ljubotina2017spin,ljubotina2019kardar,gopalakrishnan2019kinetic,de2019anomalous,gopalakrishnan2019anomalous,bulchandani2020kardar,gopalakrishnan2022anomalous}. 

\begin{figure}[tb]
    \centering
    \includegraphics[width=\linewidth]{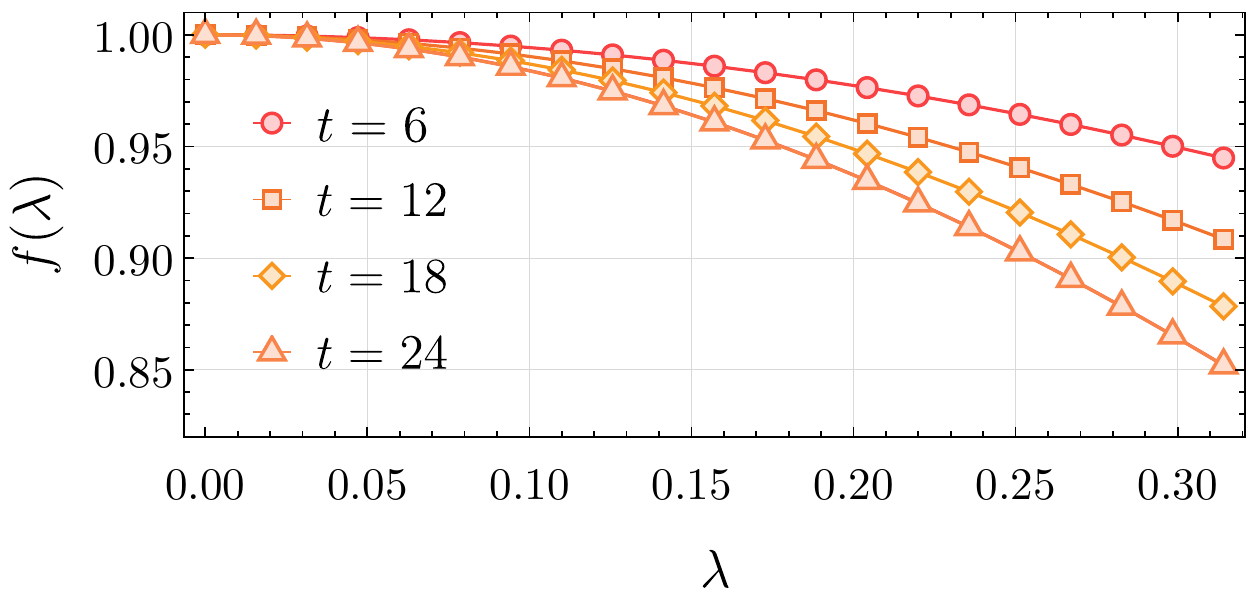}
    \caption{Moment generating function of charge transfer at equilibrium in a 46-site Trotterized Heisenberg spin chain for various depths.}
    \label{fig:MGF}
\end{figure}

\begin{figure*}[tb]
    \centering
    \includegraphics[width=\linewidth]{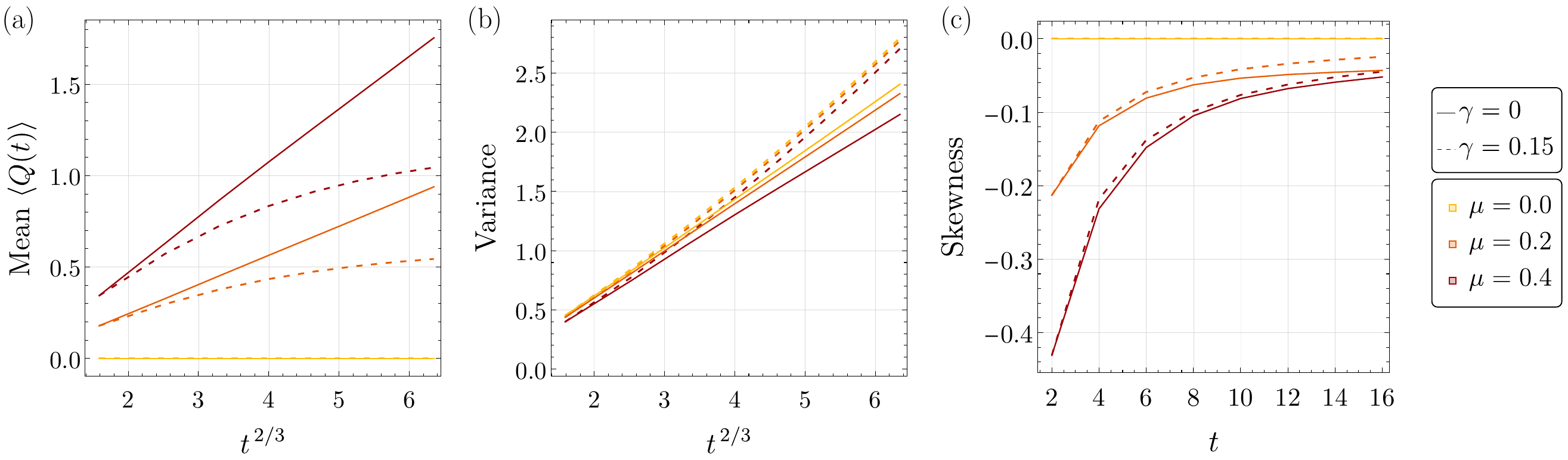}
    \caption{Mean, variance, and skewness of the transferred magnetization for 46 qubits as a function of circuit depth for time evolution under Trotterized XXZ dynamics with $\Delta=1$. We study three different domain-wall initial states described by the density matrix \eqref{eq:rho} for $\mu$\,$=$\,$0$, $0.2$, $0.4$ (distinguished by different colors) without any noise (solid lines) and under the influence of depolarizing noise with strength $\gamma=0.15$ (dashed lines). }
    \label{fig:XXZ}
\end{figure*}

In Fig.~\ref{fig:MGF}, we plot the generating function $f(\lambda)$ for $\mu$\,$=$\,$0$ obtained using the ancilla; we fit this function to a low-order polynomial near the origin in order to compute its derivatives. This approach is especially well-suited to extract the mean of the total transferred magnetization, $\langle Q (t) \rangle $, defined here as the net number of spin excitations that have crossed between the two halves of the chain by time $t$,
\begin{equation}
Q (t) = \sum_{i>N/2} \left[S^z_i(t) - S^z_i(0)\right] - \sum_{i \le N/2} \left[S^z_i(t) - S^z_i(0)\right]
\end{equation}
as well as its variance $\langle (Q (t) -\langle Q (t) \rangle)^2 \rangle$, which we can calculate up to $t$\,$=$\,$40$ using the time-evolving block decimation (TEBD) algorithm with a truncation error $< 10^{-7}$.

\textit{Adding noise.}---Another key feature of the ancilla method is that remains capable of measuring FCS in the presence of noise. 
By contrast, direct microscopy is vulnerable to any noise that violates total particle number conservation, since microscopy cannot distinguish changes in particle number due to transport across the midpoint from those due to number-changing noise deep inside the subsystem. 
In some cases, e.g., with amplitude-damping noise, the effects of noise can be detected and eliminated through postselection (albeit at the cost of exponential overhead in the subsystem size). In other cases, such as depolarizing noise, the noise cannot be corrected for even using postselection. As a result, for large subsystems in the presence of noise, microscopy can measure the FCS only on timescales when essentially no noise has acted \emph{anywhere} in the subsystem. 

The turnstile method is insensitive, by construction, to noise occurring deep in the subsystems. Hence, it faithfully captures the dynamics of charge transfer in noisy systems. When the noise is strong enough, the dynamics of charge transfer are themselves affected by the noise; the turnstile method captures this physical crossover between unitary and noisy dynamics as we now show. 

To demonstrate this point, we study transport in the Heisenberg model, in the presence of noise, for the mean, the variance, and the skewness $\langle (Q(t) -\langle Q(t) \rangle)^3 \rangle$ as a function of the circuit depth.
The third and higher moments can be obtained from the subdominant terms in the expansions for $\langle X \rangle$, $\langle Y \rangle$ and consequently, are very sensitive to the details of the fitting polynomial. It is therefore advantageous to instead construct the full probability distribution $p_Q$ by examining $\lambda\in [0, \pi]$, which avoids any fitting errors, with the tradeoff being that larger counting fields require a higher bond dimension to simulate.  

Here, we focus on depolarizing noise, described by the quantum channel %, which is particularly deleterious for experiments since it cannot be corrected for by postselecting on bitstrings with a specific number of particles. We encode the noise as a quantum channel
$
\rho$\,$\rightarrow$\,$\sum_{p=0}^3 K^{\pdagger}_p \rho K^{\dagger}_p,
$
where $K_p$ are the Kraus operators
\begin{equation*}
K^{}_0 \hspace*{-0.05cm}=\hspace*{-0.05cm} \sqrt{1-\frac{3\gamma}{4}}\, \mathds{1}, K^{}_1 \hspace*{-0.05cm}=\hspace*{-0.05cm} \sqrt{\frac{\gamma}{4}} \sigma^x, K^{}_2 \hspace*{-0.05cm}=\hspace*{-0.05cm} \sqrt{\frac{\gamma}{4}} \sigma^y, K^{}_3 \hspace*{-0.05cm}=\hspace*{-0.05cm} \sqrt{\frac{\gamma}{4}} \sigma^z.
\end{equation*}
Our results for the noisy simulations are shown in Fig.~\ref{fig:XXZ}. On timescales that are short compared to $1/\gamma$, we reproduce the noiseless behavior, with the mean and variance scaling as $t^{2/3}$~\cite{de2021stability}. On longer timescales, we find that the mean saturates and the variance begins to grow linearly. This is a consequence of the noise breaking the conservation law: on long timescales, the occupation of the central site fluctuates randomly, so the phase accumulation is random. Thus, its mean stops changing and its variance grows linearly in time. To the best of our knowledge, the turnstile method is the only way to access this crossover; while our predictions are converged in truncation error, the fact that they reproduce theoretical expectations in the large-noise limit is a helpful sanity check.

Besides the mean and the variance, the turnstile protocol also naturally facilitates access to the higher moments, which have been shown to be essential for determining the dynamical universality class in recent experiments on a superconducting quantum processor \cite{google2023}. Interstingly, Fig.~\ref{fig:XXZ}(c) indicates that the skewness is much less affected by the noise than the first two moments. These conclusions continue to hold for other types of noise, such as amplitude damping, as detailed in the Supplemental Material.

Finally, we remark on the effects of noise on the ancilla itself. In general, we can use the ancilla to measure FCS until we reach the $T_2$ time of the ancilla. In an optimally calibrated protocol, $\lambda t$ would be chosen on a uniform grid from $0$ to $1/Q_{\mathrm{typ}}$, where $Q_{\mathrm{typ}}$ is the typical scale of charge transfer over timescale $t$\,$\leq$\,$T_2$. Estimating the number of shots needed to get a precise histogram of charge transfer is a standard exercise in metrology~\cite{toth2014quantum}: this quantity scales as $N(\lambda) $\,$\sim$\,$\min(1/\lambda^2, (t/T_2)^2/\lambda^4)$. This rapid increase in the number of needed shots limits the resolution with which $Q(t)$ can be reconstructed at late times. The sensitivity limits of FCS measurements, as well as ways to improve their sensitivity using ideas from error correction~\cite{PhysRevLett.112.150802}, will be explored in future work.

\textit{Application to approximate algorithms.}---In addition to its experimental relevance, the turnstile approach has the benefit of being compatible with approximate quantum simulation algorithms. 
For purely unitary dynamics, the turnstile approach is equivalent to the introduction of a fictitious counting field~\cite{Levitov1996Electron, Nazarov1999, Belzig2001, Nazarov_2003, Levitov_2004, Bachmann_2009, Lesovik13, Tang_2014} that modifies the forward and backward evolution in different ways. When unitarity is broken, the turnstile method continues to treat the evolution of the full system (including the ancilla) as a quantum channel, but other counting-field approaches do not. This means that the turnstile method is compatible with quantum simulation algorithms that execute a quantum channel by construction. 

An example of such an algorithm, and the one that we will focus our attention on in this section, is the density matrix truncation (DMT) algorithm \cite{White2018}. By protecting carefully chosen components during a singular-value-decomposition and truncation step, DMT preserves the local structure of the system's density matrix---after a full truncation sweep over every bond, the algorithm preserves all reduced densities matrices on three neighboring sites \footnote{There are generalizations of DMT which preserve operators on $\ell$ consecutive sites \cite{Ye2020}.}. However, by treating the ancilla as a bona fide site in the chain, three-site operators on the physical sites in the neighbourhood of the ancilla are no longer preserved. We can restore the preservation of all physical three-site operators by combining the ancilla and a physical site into a super-site and modifying DMT to accommodate the enlarged site.

DMT performs similarly to other methods near equilibrium, but has shown a clear advantage far from equilibrium \cite{White2018,Ye2020}. Combining the turnstile method and the DMT algorithm should enable more efficient evaluation of nonequilibrium FCS, closing the gap to ultracold atomic and trapped ion experiments where initial states are typically far from equilibrium \cite{Cazalilla2011,Wei2022,Choi2016,Islam2015,RubioAbadal2019,Kaufman2016,Fukuhara2013,Leonard2020,Kinoshita2006,Hofferberth2007,Trotzky2012,Schreiber2015}.
In the rest of this section, we will study the nonequilibrium charge transfer statistics of a U$(1)$-charge-conserving random circuit on a spin-$1/2$ chain, using both DMT and its closest competitor, TEBD. We will focus on the N\'eel state as an initial state, from which the system undergoes a global quench. Both the global quench and the rapid entanglement generation in random-circuit evolution make this environment very challenging for existing numerical tools. 

\begin{figure}[tb]
    \centering
    \includegraphics[width=0.49\textwidth, trim= 0 0 0 1cm, clip]{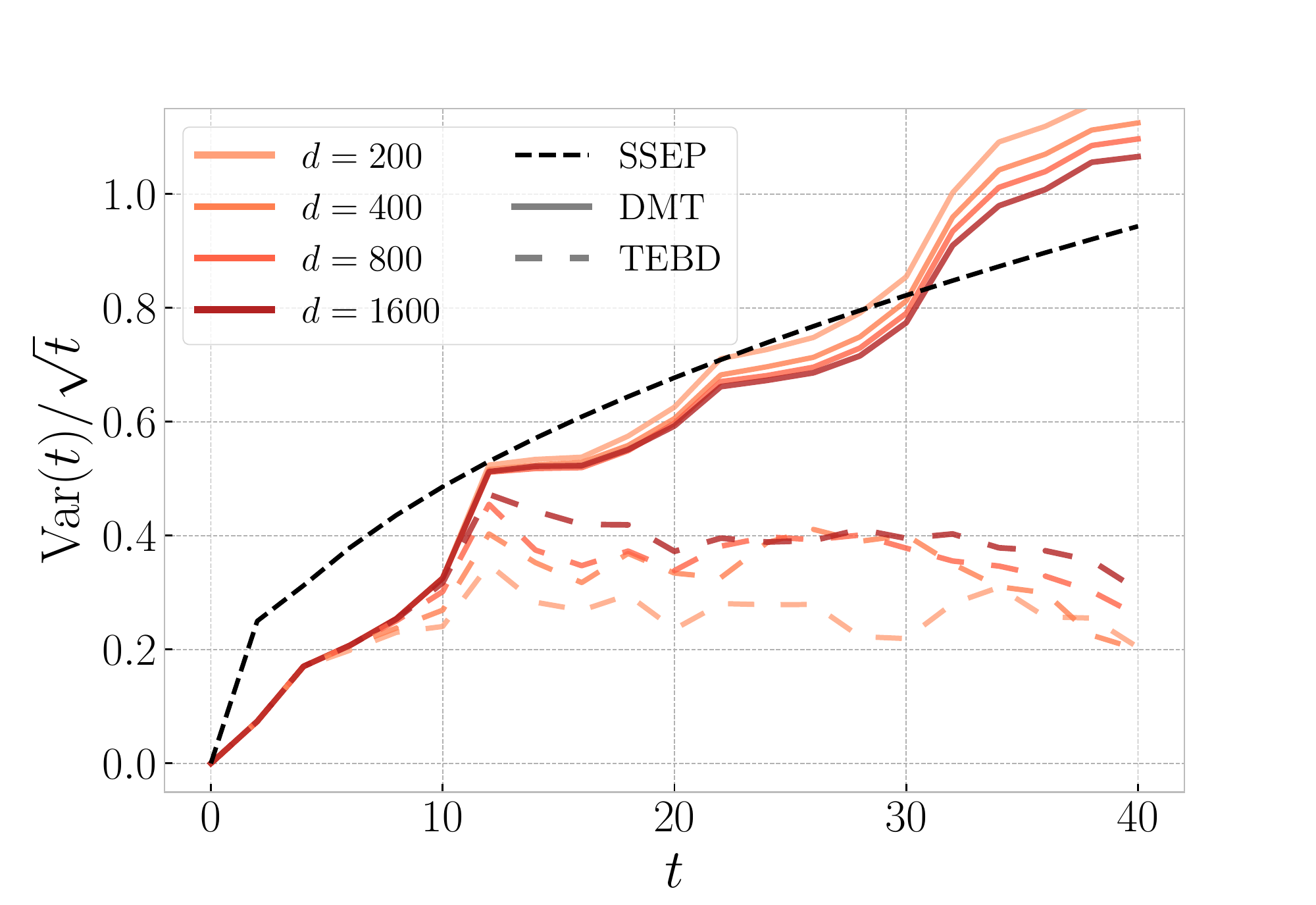}
    \caption{The variance of the charge transfer across the central bond of a random unitary circuit prepared in the N\'eel state. Both TEBD (dashed) and DMT (solid) algorithms are used with bond dimensions up to $d = 1600$. The FCS is expected to be asymptotically described by a symmetric exclusion process (SSEP) (black dashed).}
    \label{randcirc_var}
\end{figure}

We simulate the system, including the ancilla, up to time $t$\,$=$\,$40$ with a selection of counting fields and bond dimensions ($d$\,$=$\,$200,400,800,1600$) for a single circuit realization. Doing so for both TEBD and DMT, we extract the variance of the charge transfer using a low-order polynomial fitting of the cumulant generating function, $\chi(\lambda)$\,$=$\,$\log f(\lambda)$, near the origin. 
The (scaled) variance is shown in Fig.~\ref{randcirc_var}.
With regular TEBD, the data quickly becomes unconverged in bond dimension (by times $t$\,$\sim$\,$10$). DMT significantly outperforms this, remaining converged for much later times ($t$\,$\sim$\,$26$). We have verified that this separation in performance is generic for random circuits by repeating our analysis for multiple circuit realizations, as well as for a different nonequilibrium state (a fully polarized domain-wall state), which can be found in the Supplemental Material~\footnote{If one is restricted to pure states, TEBD (as well as exact statevector simulations) are competitive with DMT. Unlike these, however, DMT can be adapted to the experimentally realistic case where the system is in a mixed state.}. 

In addition to improved convergence characteristics, the DMT results are consistent with the FCS in a symmetric simple exclusion process, which is expected to describe the asymptotic FCS of U$(1)$-charge-conserving random unitary circuits \cite{McCulloch2023}. The diffusivity $D$\,$=$\,$1$ and the susceptibility $\chi$ are the same for a symmetric exclusion process and the random circuits considered here. Therefore, the dynamics of charge, including at the level of fluctuations, is given by the same fluctuating hydrodynamic theory \cite{Bertini_2015}. It follows that a symmetric exclusion process and the random circuit considered here should share the same long-time charge transfer statistics. This is borne out by Fig.~\ref{randcirc_var}, which overlays the charge transfer variance for the exclusion process initialized with a N\'eel state over the random circuit results.

We now briefly compare the performance of DMT using the turnstile with that of other numerical approaches (beyond the comparison with mixed-state TEBD above). For pure-state evolution, DMT can be compared with either direct matrix-vector multiplication or pure-state TEBD~\footnote{We remark that either the turnstile or the related counting-field method is needed with TEBD in any case, to allow one to treat initial states that are not number-sharp.}. For comparable resources, those methods are accurate until times in the range $t \leq 24$--$32$. For mixed states, exact density-matrix evolution and TEBD with quantum trajectories are limited to much shorter times $t \leq 16$ (in the latter case, because of sampling overhead)~\cite{daley2014quantum}. In the equilibrium limit, as noted above, the performance of DMT is comparable to that of mixed-state TEBD. Thus, in the pure-state and equilibrium limits, DMT is competitive with algorithms that exploit the special structure present in those limits. Away from those limits---for evolving generic nonequilibrium mixed states---DMT outperforms these alternative methods, by approximately a factor of two. 

\textit{Discussion and outlook.}---To date, computing FCS has been a challenging task---both experimentally and theoretically---as it requires knowledge of arbitrary-order correlations. 
In this work, we overcome these obstacles and present a new method for measuring FCS, based on a quantum turnstile, that is robust against noise and only involves local operators, thus allowing for its easy implementation in near-term experimental platforms. We showed the versatility of this approach in the context of studying the FCS of magnetization transfer in a Heisenberg spin chain. Our results in this regard directly apply to the recent experiments on superconducting qubit arrays \cite{google2023}, which rely on the higher moments, such as the skewness and kurtosis, to determine the dynamical universality class. We also illustrated how the quantum turnstile protocol  can be paired with tensor-network algorithms to study numerically demanding transport problems such as charge transfer in random circuits. Our results pave the way for using FCS to resolve the underlying dynamics of a range of many-body systems that are realizable in noisy quantum simulators (e.g., disordered systems). The turnstile method can be further adapted to explore higher-order temporal fluctuations of more general (i.e., nonconserved) operators in many-body systems; the nature of these fluctuations is largely an open question, although exact predictions have been derived for certain solvable models~\cite{klobas2020matrix}.

\begin{acknowledgments}
	We thank Trond Andersen, Juan P. Garrahan, Toma\v{z} Prosen, Eliott Rosenberg, and Pedram Roushan for many insightful discussions. We acknowledge support from the Princeton Quantum Initiative Fellowship (R.S.), the National Science Foundation through grants DMR-2103938 (E.M., S.G.) and DMR-2104141 (E.M., R.V.), the US Department of Energy, Office of Science, Basic Energy Sciences, under Early Career Award No. DE-SC0021111 (V.K.), the Alfred P. Sloan Foundation through a Sloan Research Fellowship (R.V., V.K.), and the Packard Foundation through a Packard Fellowship (V.K.). Some of the simulations presented in this article were performed on computational resources managed and supported by Princeton Research Computing, a consortium of groups including the Princeton Institute for Computational Science and Engineering (PICSciE) and the Office of Information Technology's High Performance Computing Center and Visualization Laboratory at Princeton University.
\end{acknowledgments}

\iffalse
The algorithm is a modification to the truncation step in the time-evolving block-decimiation algorithm, which we will briefly review now. We begin by Schmidt decomposing a density matrix $\rho$ at bond $\{j,j+1\}$,
\begin{equation}
\rho = \sum_{a=0}^{\chi-1} x_{L a}[j] s_a[j] x_{R a}[j],
\end{equation}
where the left and right Schmidt vectors are given (in the basis of Pauli matrices $\sigma^{\mu}\in\{\mathds{1},\sigma^x,\sigma^y,\sigma^z\}$, with $\mu = 0,\cdots,3$ ) by
\begin{align}
    x_{L a}[j] &= \sum_{\mu_1\cdots \mu_j} \left[ A_{(1)}^{\mu_1}\cdots A_{(j)}^{\mu_j}\right]_{a} \sigma^{\mu_1}\cdots \sigma^{\mu_j}\\
    x_{R a}[j] &= \sum_{\mu_{j+1}\cdots \mu_{\textsc{l}}} \left[ A_{(j+1)}^{\mu_{j+1}}\cdots A_{(L)}^{\mu_{\textsc{l}}}\right]_{a} \sigma^{\mu_{j+1}}\cdots \sigma^{\mu_{\textsc{l}}}.
\end{align}
The novel step in DMT is to then rotate these Schmidt vectors $x_{\eta}\to y_{\eta}$ ($\eta = L,R$),
\begin{equation}
    y_{\eta a} = \sum_{b=0}^{\chi-1} x_{\eta b} Q^*_{\eta b a},
\end{equation}
where the rotation $Q$ is defined by the QR factorization
\begin{equation}
    Q_{\eta a b}R_{\eta b}^{\mu} = \Tr(x_{\eta a}\sigma_j^{\mu}).
\end{equation}
\fi

\bibliographystyle{apsrev4-2_custom}
\bibliography{refs.bib}

%apsrev4-2.bst 2019-01-14 (MD) hand-edited version of apsrev4-1.bst
%Control: key (0)
%Control: author (72) initials jnrlst
%Control: editor formatted (1) identically to author
%Control: production of article title (1) required
%Control: page (0) single
%Control: year (1) truncated
%Control: production of eprint (0) enabled
\begin{thebibliography}{83}%
\makeatletter
\providecommand \@ifxundefined [1]{%
 \@ifx{#1\undefined}
}%
\providecommand \@ifnum [1]{%
 \ifnum #1\expandafter \@firstoftwo
 \else \expandafter \@secondoftwo
 \fi
}%
\providecommand \@ifx [1]{%
 \ifx #1\expandafter \@firstoftwo
 \else \expandafter \@secondoftwo
 \fi
}%
\providecommand \natexlab [1]{#1}%
\providecommand \enquote  [1]{``#1''}%
\providecommand \bibnamefont  [1]{#1}%
\providecommand \bibfnamefont [1]{#1}%
\providecommand \citenamefont [1]{#1}%
\providecommand \href@noop [0]{\@secondoftwo}%
\providecommand \href [0]{\begingroup \@sanitize@url \@href}%
\providecommand \@href[1]{\@@startlink{#1}\@@href}%
\providecommand \@@href[1]{\endgroup#1\@@endlink}%
\providecommand \@sanitize@url [0]{\catcode `\\12\catcode `\$12\catcode
  `\&12\catcode `\#12\catcode `\^12\catcode `\_12\catcode `\%12\relax}%
\providecommand \@@startlink[1]{}%
\providecommand \@@endlink[0]{}%
\providecommand \url  [0]{\begingroup\@sanitize@url \@url }%
\providecommand \@url [1]{\endgroup\@href {#1}{\urlprefix }}%
\providecommand \urlprefix  [0]{URL }%
\providecommand \Eprint [0]{\href }%
\providecommand \doibase [0]{https://doi.org/}%
\providecommand \selectlanguage [0]{\@gobble}%
\providecommand \bibinfo  [0]{\@secondoftwo}%
\providecommand \bibfield  [0]{\@secondoftwo}%
\providecommand \translation [1]{[#1]}%
\providecommand \BibitemOpen [0]{}%
\providecommand \bibitemStop [0]{}%
\providecommand \bibitemNoStop [0]{.\EOS\space}%
\providecommand \EOS [0]{\spacefactor3000\relax}%
\providecommand \BibitemShut  [1]{\csname bibitem#1\endcsname}%
\let\auto@bib@innerbib\@empty
%</preamble>
\bibitem [{\citenamefont {Wan}\ and\ \citenamefont
  {Armitage}(2019)}]{PhysRevLett.122.257401}%
  \BibitemOpen
  \bibfield  {author} {\bibinfo {author} {\bibfnamefont {Y.}~\bibnamefont
  {Wan}}\ and\ \bibinfo {author} {\bibfnamefont {N.~P.}\ \bibnamefont
  {Armitage}},\ }\bibfield  {title} {\bibinfo {title} {{Resolving Continua of
  Fractional Excitations by Spinon Echo in THz 2D Coherent Spectroscopy}},\
  }\href {https://doi.org/10.1103/PhysRevLett.122.257401} {\bibfield  {journal}
  {\bibinfo  {journal} {Phys. Rev. Lett.}\ }\textbf {\bibinfo {volume} {122}},\
  \bibinfo {pages} {257401} (\bibinfo {year} {2019})}\BibitemShut {NoStop}%
\bibitem [{\citenamefont {Parker}\ \emph {et~al.}(2019)\citenamefont {Parker},
  \citenamefont {Morimoto}, \citenamefont {Orenstein},\ and\ \citenamefont
  {Moore}}]{PhysRevB.99.045121}%
  \BibitemOpen
  \bibfield  {author} {\bibinfo {author} {\bibfnamefont {D.~E.}\ \bibnamefont
  {Parker}}, \bibinfo {author} {\bibfnamefont {T.}~\bibnamefont {Morimoto}},
  \bibinfo {author} {\bibfnamefont {J.}~\bibnamefont {Orenstein}},\ and\
  \bibinfo {author} {\bibfnamefont {J.~E.}\ \bibnamefont {Moore}},\ }\bibfield
  {title} {\bibinfo {title} {{Diagrammatic approach to nonlinear optical
  response with application to Weyl semimetals}},\ }\href
  {https://doi.org/10.1103/PhysRevB.99.045121} {\bibfield  {journal} {\bibinfo
  {journal} {Phys. Rev. B}\ }\textbf {\bibinfo {volume} {99}},\ \bibinfo
  {pages} {045121} (\bibinfo {year} {2019})}\BibitemShut {NoStop}%
\bibitem [{\citenamefont {Watanabe}\ and\ \citenamefont
  {Oshikawa}(2020)}]{PhysRevB.102.165137}%
  \BibitemOpen
  \bibfield  {author} {\bibinfo {author} {\bibfnamefont {H.}~\bibnamefont
  {Watanabe}}\ and\ \bibinfo {author} {\bibfnamefont {M.}~\bibnamefont
  {Oshikawa}},\ }\bibfield  {title} {\bibinfo {title} {{Generalized $f$-sum
  rules and Kohn formulas on nonlinear conductivities}},\ }\href
  {https://doi.org/10.1103/PhysRevB.102.165137} {\bibfield  {journal} {\bibinfo
   {journal} {Phys. Rev. B}\ }\textbf {\bibinfo {volume} {102}},\ \bibinfo
  {pages} {165137} (\bibinfo {year} {2020})}\BibitemShut {NoStop}%
\bibitem [{\citenamefont {Mahmood}\ \emph {et~al.}(2021)\citenamefont
  {Mahmood}, \citenamefont {Chaudhuri}, \citenamefont {Gopalakrishnan},
  \citenamefont {Nandkishore},\ and\ \citenamefont
  {Armitage}}]{mahmood2021observation}%
  \BibitemOpen
  \bibfield  {author} {\bibinfo {author} {\bibfnamefont {F.}~\bibnamefont
  {Mahmood}}, \bibinfo {author} {\bibfnamefont {D.}~\bibnamefont {Chaudhuri}},
  \bibinfo {author} {\bibfnamefont {S.}~\bibnamefont {Gopalakrishnan}},
  \bibinfo {author} {\bibfnamefont {R.}~\bibnamefont {Nandkishore}},\ and\
  \bibinfo {author} {\bibfnamefont {N.~P.}\ \bibnamefont {Armitage}},\
  }\bibfield  {title} {\bibinfo {title} {{Observation of a marginal Fermi
  glass}},\ }\href {https://doi.org/10.1038/s41567-020-01149-0} {\bibfield
  {journal} {\bibinfo  {journal} {Nat. Phys.}\ }\textbf {\bibinfo {volume}
  {17}},\ \bibinfo {pages} {627} (\bibinfo {year} {2021})}\BibitemShut
  {NoStop}%
\bibitem [{\citenamefont {Fava}\ \emph {et~al.}(2021)\citenamefont {Fava},
  \citenamefont {Biswas}, \citenamefont {Gopalakrishnan}, \citenamefont
  {Vasseur},\ and\ \citenamefont {Parameswaran}}]{fava2021hydrodynamic}%
  \BibitemOpen
  \bibfield  {author} {\bibinfo {author} {\bibfnamefont {M.}~\bibnamefont
  {Fava}}, \bibinfo {author} {\bibfnamefont {S.}~\bibnamefont {Biswas}},
  \bibinfo {author} {\bibfnamefont {S.}~\bibnamefont {Gopalakrishnan}},
  \bibinfo {author} {\bibfnamefont {R.}~\bibnamefont {Vasseur}},\ and\ \bibinfo
  {author} {\bibfnamefont {S.}~\bibnamefont {Parameswaran}},\ }\bibfield
  {title} {\bibinfo {title} {Hydrodynamic nonlinear response of interacting
  integrable systems},\ }\href {https://doi.org/10.1073/pnas.2106945118}
  {\bibfield  {journal} {\bibinfo  {journal} {Proc. Natl. Acad. Sci. U.S.A.}\
  }\textbf {\bibinfo {volume} {118}},\ \bibinfo {pages} {e2106945118} (\bibinfo
  {year} {2021})}\BibitemShut {NoStop}%
\bibitem [{\citenamefont {Nandkishore}\ \emph {et~al.}(2021)\citenamefont
  {Nandkishore}, \citenamefont {Choi},\ and\ \citenamefont
  {Kim}}]{PhysRevResearch.3.013254}%
  \BibitemOpen
  \bibfield  {author} {\bibinfo {author} {\bibfnamefont {R.~M.}\ \bibnamefont
  {Nandkishore}}, \bibinfo {author} {\bibfnamefont {W.}~\bibnamefont {Choi}},\
  and\ \bibinfo {author} {\bibfnamefont {Y.~B.}\ \bibnamefont {Kim}},\
  }\bibfield  {title} {\bibinfo {title} {Spectroscopic fingerprints of gapped
  quantum spin liquids, both conventional and fractonic},\ }\href
  {https://doi.org/10.1103/PhysRevResearch.3.013254} {\bibfield  {journal}
  {\bibinfo  {journal} {Phys. Rev. Res.}\ }\textbf {\bibinfo {volume} {3}},\
  \bibinfo {pages} {013254} (\bibinfo {year} {2021})}\BibitemShut {NoStop}%
\bibitem [{\citenamefont {Tanikawa}\ \emph {et~al.}(2021)\citenamefont
  {Tanikawa}, \citenamefont {Takasan},\ and\ \citenamefont
  {Katsura}}]{PhysRevB.103.L201120}%
  \BibitemOpen
  \bibfield  {author} {\bibinfo {author} {\bibfnamefont {Y.}~\bibnamefont
  {Tanikawa}}, \bibinfo {author} {\bibfnamefont {K.}~\bibnamefont {Takasan}},\
  and\ \bibinfo {author} {\bibfnamefont {H.}~\bibnamefont {Katsura}},\
  }\bibfield  {title} {\bibinfo {title} {{Exact results for nonlinear Drude
  weights in the spin-$\frac{1}{2}$ XXZ chain}},\ }\href
  {https://doi.org/10.1103/PhysRevB.103.L201120} {\bibfield  {journal}
  {\bibinfo  {journal} {Phys. Rev. B}\ }\textbf {\bibinfo {volume} {103}},\
  \bibinfo {pages} {L201120} (\bibinfo {year} {2021})}\BibitemShut {NoStop}%
\bibitem [{\citenamefont {Li}\ \emph {et~al.}(2022)\citenamefont {Li},
  \citenamefont {Ning}, \citenamefont {Mehio}, \citenamefont {Zhao},
  \citenamefont {Lee}, \citenamefont {Kim}, \citenamefont {Nakamura},
  \citenamefont {Maeno}, \citenamefont {Cao},\ and\ \citenamefont
  {Hsieh}}]{PhysRevLett.128.187402}%
  \BibitemOpen
  \bibfield  {author} {\bibinfo {author} {\bibfnamefont {X.}~\bibnamefont
  {Li}}, \bibinfo {author} {\bibfnamefont {H.}~\bibnamefont {Ning}}, \bibinfo
  {author} {\bibfnamefont {O.}~\bibnamefont {Mehio}}, \bibinfo {author}
  {\bibfnamefont {H.}~\bibnamefont {Zhao}}, \bibinfo {author} {\bibfnamefont
  {M.-C.}\ \bibnamefont {Lee}}, \bibinfo {author} {\bibfnamefont
  {K.}~\bibnamefont {Kim}}, \bibinfo {author} {\bibfnamefont {F.}~\bibnamefont
  {Nakamura}}, \bibinfo {author} {\bibfnamefont {Y.}~\bibnamefont {Maeno}},
  \bibinfo {author} {\bibfnamefont {G.}~\bibnamefont {Cao}},\ and\ \bibinfo
  {author} {\bibfnamefont {D.}~\bibnamefont {Hsieh}},\ }\bibfield  {title}
  {\bibinfo {title} {{Keldysh Space Control of Charge Dynamics in a Strongly
  Driven Mott Insulator}},\ }\href
  {https://doi.org/10.1103/PhysRevLett.128.187402} {\bibfield  {journal}
  {\bibinfo  {journal} {Phys. Rev. Lett.}\ }\textbf {\bibinfo {volume} {128}},\
  \bibinfo {pages} {187402} (\bibinfo {year} {2022})}\BibitemShut {NoStop}%
\bibitem [{\citenamefont {Kane}(2022)}]{PhysRevLett.128.076801}%
  \BibitemOpen
  \bibfield  {author} {\bibinfo {author} {\bibfnamefont {C.~L.}\ \bibnamefont
  {Kane}},\ }\bibfield  {title} {\bibinfo {title} {Quantized nonlinear
  conductance in ballistic metals},\ }\href
  {https://doi.org/10.1103/PhysRevLett.128.076801} {\bibfield  {journal}
  {\bibinfo  {journal} {Phys. Rev. Lett.}\ }\textbf {\bibinfo {volume} {128}},\
  \bibinfo {pages} {076801} (\bibinfo {year} {2022})}\BibitemShut {NoStop}%
\bibitem [{\citenamefont {McGinley}\ \emph {et~al.}(2022)\citenamefont
  {McGinley}, \citenamefont {Fava},\ and\ \citenamefont
  {Parameswaran}}]{mcginley2022signatures}%
  \BibitemOpen
  \bibfield  {author} {\bibinfo {author} {\bibfnamefont {M.}~\bibnamefont
  {McGinley}}, \bibinfo {author} {\bibfnamefont {M.}~\bibnamefont {Fava}},\
  and\ \bibinfo {author} {\bibfnamefont {S.~A.}\ \bibnamefont {Parameswaran}},\
  }\bibfield  {title} {\bibinfo {title} {Signatures of fractional statistics in
  nonlinear pump-probe spectroscopy},\ }\href
  {https://arxiv.org/abs/2210.16249} {\bibfield  {journal} {\bibinfo  {journal}
  {arXiv preprint arXiv:2210.16249}\ } (\bibinfo {year} {2022})}\BibitemShut
  {NoStop}%
\bibitem [{\citenamefont {Blanter}\ and\ \citenamefont
  {B{\"u}ttiker}(2000)}]{blanter2000shot}%
  \BibitemOpen
  \bibfield  {author} {\bibinfo {author} {\bibfnamefont {Y.~M.}\ \bibnamefont
  {Blanter}}\ and\ \bibinfo {author} {\bibfnamefont {M.}~\bibnamefont
  {B{\"u}ttiker}},\ }\bibfield  {title} {\bibinfo {title} {Shot noise in
  mesoscopic conductors},\ }\href
  {https://doi.org/10.1016/S0370-1573(99)00123-4} {\bibfield  {journal}
  {\bibinfo  {journal} {Phys. Rep.}\ }\textbf {\bibinfo {volume} {336}},\
  \bibinfo {pages} {1} (\bibinfo {year} {2000})}\BibitemShut {NoStop}%
\bibitem [{\citenamefont {Kolkowitz}\ \emph {et~al.}(2015)\citenamefont
  {Kolkowitz}, \citenamefont {Safira}, \citenamefont {High}, \citenamefont
  {Devlin}, \citenamefont {Choi}, \citenamefont {Unterreithmeier},
  \citenamefont {Patterson}, \citenamefont {Zibrov}, \citenamefont
  {Manucharyan}, \citenamefont {Park},\ and\ \citenamefont
  {Lukin}}]{kolkowitz2015probing}%
  \BibitemOpen
  \bibfield  {author} {\bibinfo {author} {\bibfnamefont {S.}~\bibnamefont
  {Kolkowitz}}, \bibinfo {author} {\bibfnamefont {A.}~\bibnamefont {Safira}},
  \bibinfo {author} {\bibfnamefont {A.}~\bibnamefont {High}}, \bibinfo {author}
  {\bibfnamefont {R.}~\bibnamefont {Devlin}}, \bibinfo {author} {\bibfnamefont
  {S.}~\bibnamefont {Choi}}, \bibinfo {author} {\bibfnamefont {Q.}~\bibnamefont
  {Unterreithmeier}}, \bibinfo {author} {\bibfnamefont {D.}~\bibnamefont
  {Patterson}}, \bibinfo {author} {\bibfnamefont {A.~S.}\ \bibnamefont
  {Zibrov}}, \bibinfo {author} {\bibfnamefont {V.~E.}\ \bibnamefont
  {Manucharyan}}, \bibinfo {author} {\bibfnamefont {H.}~\bibnamefont {Park}},\
  and\ \bibinfo {author} {\bibfnamefont {M.~D.}\ \bibnamefont {Lukin}},\
  }\bibfield  {title} {\bibinfo {title} {{Probing Johnson noise and ballistic
  transport in normal metals with a single-spin qubit}},\ }\href
  {https://doi.org/10.1126/science.aaa4298} {\bibfield  {journal} {\bibinfo
  {journal} {Science}\ }\textbf {\bibinfo {volume} {347}},\ \bibinfo {pages}
  {1129} (\bibinfo {year} {2015})}\BibitemShut {NoStop}%
\bibitem [{\citenamefont {Sinitsyn}\ and\ \citenamefont
  {Pershin}(2016)}]{sinitsyn2016theory}%
  \BibitemOpen
  \bibfield  {author} {\bibinfo {author} {\bibfnamefont {N.~A.}\ \bibnamefont
  {Sinitsyn}}\ and\ \bibinfo {author} {\bibfnamefont {Y.~V.}\ \bibnamefont
  {Pershin}},\ }\bibfield  {title} {\bibinfo {title} {The theory of spin noise
  spectroscopy: a review},\ }\href
  {https://doi.org/10.1088/0034-4885/79/10/106501} {\bibfield  {journal}
  {\bibinfo  {journal} {Rep. Prog. Phys.}\ }\textbf {\bibinfo {volume} {79}},\
  \bibinfo {pages} {106501} (\bibinfo {year} {2016})}\BibitemShut {NoStop}%
\bibitem [{\citenamefont {Rovny}\ \emph {et~al.}(2022)\citenamefont {Rovny},
  \citenamefont {Yuan}, \citenamefont {Fitzpatrick}, \citenamefont {Abdalla},
  \citenamefont {Futamura}, \citenamefont {Fox}, \citenamefont {Cambria},
  \citenamefont {Kolkowitz},\ and\ \citenamefont
  {de~Leon}}]{rovny2022nanoscale}%
  \BibitemOpen
  \bibfield  {author} {\bibinfo {author} {\bibfnamefont {J.}~\bibnamefont
  {Rovny}}, \bibinfo {author} {\bibfnamefont {Z.}~\bibnamefont {Yuan}},
  \bibinfo {author} {\bibfnamefont {M.}~\bibnamefont {Fitzpatrick}}, \bibinfo
  {author} {\bibfnamefont {A.~I.}\ \bibnamefont {Abdalla}}, \bibinfo {author}
  {\bibfnamefont {L.}~\bibnamefont {Futamura}}, \bibinfo {author}
  {\bibfnamefont {C.}~\bibnamefont {Fox}}, \bibinfo {author} {\bibfnamefont
  {M.~C.}\ \bibnamefont {Cambria}}, \bibinfo {author} {\bibfnamefont
  {S.}~\bibnamefont {Kolkowitz}},\ and\ \bibinfo {author} {\bibfnamefont
  {N.~P.}\ \bibnamefont {de~Leon}},\ }\bibfield  {title} {\bibinfo {title}
  {Nanoscale covariance magnetometry with diamond quantum sensors},\ }\href
  {https://doi.org/10.1126/science.ade9858} {\bibfield  {journal} {\bibinfo
  {journal} {Science}\ }\textbf {\bibinfo {volume} {378}},\ \bibinfo {pages}
  {1301} (\bibinfo {year} {2022})}\BibitemShut {NoStop}%
\bibitem [{\citenamefont {{Levitov}}\ and\ \citenamefont
  {{Lesovik}}(1993)}]{Levitov1993}%
  \BibitemOpen
  \bibfield  {author} {\bibinfo {author} {\bibfnamefont {L.~S.}\ \bibnamefont
  {{Levitov}}}\ and\ \bibinfo {author} {\bibfnamefont {G.~B.}\ \bibnamefont
  {{Lesovik}}},\ }\bibfield  {title} {\bibinfo {title} {{Charge distribution in
  quantum shot noise}},\ }\href
  {http://jetpletters.ru/ps/1186/article_17907.pdf} {\bibfield  {journal}
  {\bibinfo  {journal} {JETP Lett.}\ }\textbf {\bibinfo {volume} {58}},\
  \bibinfo {pages} {230} (\bibinfo {year} {1993})}\BibitemShut {NoStop}%
\bibitem [{\citenamefont {Levitov}\ \emph {et~al.}(1996)\citenamefont
  {Levitov}, \citenamefont {Lee},\ and\ \citenamefont
  {Lesovik}}]{Levitov1996Electron}%
  \BibitemOpen
  \bibfield  {author} {\bibinfo {author} {\bibfnamefont {L.~S.}\ \bibnamefont
  {Levitov}}, \bibinfo {author} {\bibfnamefont {H.}~\bibnamefont {Lee}},\ and\
  \bibinfo {author} {\bibfnamefont {G.~B.}\ \bibnamefont {Lesovik}},\
  }\bibfield  {title} {\bibinfo {title} {Electron counting statistics and
  coherent states of electric current},\ }\href
  {https://doi.org/10.1063/1.531672} {\bibfield  {journal} {\bibinfo  {journal}
  {J. Math. Phys.}\ }\textbf {\bibinfo {volume} {37}},\ \bibinfo {pages} {4845}
  (\bibinfo {year} {1996})}\BibitemShut {NoStop}%
\bibitem [{\citenamefont {{Belzig}}\ and\ \citenamefont
  {{Nazarov}}(2001)}]{Belzig2001}%
  \BibitemOpen
  \bibfield  {author} {\bibinfo {author} {\bibfnamefont {W.}~\bibnamefont
  {{Belzig}}}\ and\ \bibinfo {author} {\bibfnamefont {Y.~V.}\ \bibnamefont
  {{Nazarov}}},\ }\bibfield  {title} {\bibinfo {title} {{Full Counting
  Statistics of Electron Transfer between Superconductors}},\ }\href
  {https://doi.org/10.1103/PhysRevLett.87.197006} {\bibfield  {journal}
  {\bibinfo  {journal} {\prl}\ }\textbf {\bibinfo {volume} {87}},\ \bibinfo
  {eid} {197006} (\bibinfo {year} {2001})}\BibitemShut {NoStop}%
\bibitem [{\citenamefont {{B{\"o}rlin}}\ \emph {et~al.}(2002)\citenamefont
  {{B{\"o}rlin}}, \citenamefont {{Belzig}},\ and\ \citenamefont
  {{Bruder}}}]{Belzig2002}%
  \BibitemOpen
  \bibfield  {author} {\bibinfo {author} {\bibfnamefont {J.}~\bibnamefont
  {{B{\"o}rlin}}}, \bibinfo {author} {\bibfnamefont {W.}~\bibnamefont
  {{Belzig}}},\ and\ \bibinfo {author} {\bibfnamefont {C.}~\bibnamefont
  {{Bruder}}},\ }\bibfield  {title} {\bibinfo {title} {{Full Counting
  Statistics of a Superconducting Beam Splitter}},\ }\href
  {https://doi.org/10.1103/PhysRevLett.88.197001} {\bibfield  {journal}
  {\bibinfo  {journal} {\prl}\ }\textbf {\bibinfo {volume} {88}},\ \bibinfo
  {eid} {197001} (\bibinfo {year} {2002})}\BibitemShut {NoStop}%
\bibitem [{\citenamefont {{Levitov}}\ and\ \citenamefont
  {{Reznikov}}(2004)}]{Levitov_2004}%
  \BibitemOpen
  \bibfield  {author} {\bibinfo {author} {\bibfnamefont {L.~S.}\ \bibnamefont
  {{Levitov}}}\ and\ \bibinfo {author} {\bibfnamefont {M.}~\bibnamefont
  {{Reznikov}}},\ }\bibfield  {title} {\bibinfo {title} {{Counting statistics
  of tunneling current}},\ }\href {https://doi.org/10.1103/PhysRevB.70.115305}
  {\bibfield  {journal} {\bibinfo  {journal} {\prb}\ }\textbf {\bibinfo
  {volume} {70}},\ \bibinfo {eid} {115305} (\bibinfo {year}
  {2004})}\BibitemShut {NoStop}%
\bibitem [{\citenamefont {{Gogolin}}\ and\ \citenamefont
  {{Komnik}}(2006)}]{Gogolin_2006}%
  \BibitemOpen
  \bibfield  {author} {\bibinfo {author} {\bibfnamefont {A.~O.}\ \bibnamefont
  {{Gogolin}}}\ and\ \bibinfo {author} {\bibfnamefont {A.}~\bibnamefont
  {{Komnik}}},\ }\bibfield  {title} {\bibinfo {title} {{Towards full counting
  statistics for the Anderson impurity model}},\ }\href
  {https://doi.org/10.1103/PhysRevB.73.195301} {\bibfield  {journal} {\bibinfo
  {journal} {\prb}\ }\textbf {\bibinfo {volume} {73}},\ \bibinfo {eid} {195301}
  (\bibinfo {year} {2006})}\BibitemShut {NoStop}%
\bibitem [{\citenamefont {Levitov}(2003)}]{Levitov2003}%
  \BibitemOpen
  \bibfield  {author} {\bibinfo {author} {\bibfnamefont {L.~S.}\ \bibnamefont
  {Levitov}},\ }\bibinfo {title} {The statistical theory of mesoscopic noise},\
  in\ \href {https://doi.org/10.1007/978-94-010-0089-5_18} {\emph {\bibinfo
  {booktitle} {Quantum Noise in Mesoscopic Physics}}},\ \bibinfo {editor}
  {edited by\ \bibinfo {editor} {\bibfnamefont {Y.~V.}\ \bibnamefont
  {Nazarov}}}\ (\bibinfo  {publisher} {Springer Netherlands},\ \bibinfo
  {address} {Dordrecht},\ \bibinfo {year} {2003})\ pp.\ \bibinfo {pages}
  {373--396}\BibitemShut {NoStop}%
\bibitem [{\citenamefont {{Bakr}}\ \emph {et~al.}(2009)\citenamefont {{Bakr}},
  \citenamefont {{Gillen}}, \citenamefont {{Peng}}, \citenamefont
  {{F{\"o}lling}},\ and\ \citenamefont {{Greiner}}}]{Bakr_2009}%
  \BibitemOpen
  \bibfield  {author} {\bibinfo {author} {\bibfnamefont {W.~S.}\ \bibnamefont
  {{Bakr}}}, \bibinfo {author} {\bibfnamefont {J.~I.}\ \bibnamefont
  {{Gillen}}}, \bibinfo {author} {\bibfnamefont {A.}~\bibnamefont {{Peng}}},
  \bibinfo {author} {\bibfnamefont {S.}~\bibnamefont {{F{\"o}lling}}},\ and\
  \bibinfo {author} {\bibfnamefont {M.}~\bibnamefont {{Greiner}}},\ }\bibfield
  {title} {\bibinfo {title} {{A quantum gas microscope for detecting single
  atoms in a Hubbard-regime optical lattice}},\ }\href
  {https://doi.org/10.1038/nature08482} {\bibfield  {journal} {\bibinfo
  {journal} {Nature}\ }\textbf {\bibinfo {volume} {462}},\ \bibinfo {pages}
  {74} (\bibinfo {year} {2009})}\BibitemShut {NoStop}%
\bibitem [{\citenamefont {{Sherson}}\ \emph {et~al.}(2010)\citenamefont
  {{Sherson}}, \citenamefont {{Weitenberg}}, \citenamefont {{Endres}},
  \citenamefont {{Cheneau}}, \citenamefont {{Bloch}},\ and\ \citenamefont
  {{Kuhr}}}]{Sherson2010}%
  \BibitemOpen
  \bibfield  {author} {\bibinfo {author} {\bibfnamefont {J.~F.}\ \bibnamefont
  {{Sherson}}}, \bibinfo {author} {\bibfnamefont {C.}~\bibnamefont
  {{Weitenberg}}}, \bibinfo {author} {\bibfnamefont {M.}~\bibnamefont
  {{Endres}}}, \bibinfo {author} {\bibfnamefont {M.}~\bibnamefont {{Cheneau}}},
  \bibinfo {author} {\bibfnamefont {I.}~\bibnamefont {{Bloch}}},\ and\ \bibinfo
  {author} {\bibfnamefont {S.}~\bibnamefont {{Kuhr}}},\ }\bibfield  {title}
  {\bibinfo {title} {{Single-atom-resolved fluorescence imaging of an atomic
  Mott insulator}},\ }\href {https://doi.org/10.1038/nature09378} {\bibfield
  {journal} {\bibinfo  {journal} {Nature}\ }\textbf {\bibinfo {volume} {467}},\
  \bibinfo {pages} {68} (\bibinfo {year} {2010})}\BibitemShut {NoStop}%
\bibitem [{\citenamefont {{Bloch}}\ \emph {et~al.}(2012)\citenamefont
  {{Bloch}}, \citenamefont {{Dalibard}},\ and\ \citenamefont
  {{Nascimb{\`e}ne}}}]{bloch2012}%
  \BibitemOpen
  \bibfield  {author} {\bibinfo {author} {\bibfnamefont {I.}~\bibnamefont
  {{Bloch}}}, \bibinfo {author} {\bibfnamefont {J.}~\bibnamefont
  {{Dalibard}}},\ and\ \bibinfo {author} {\bibfnamefont {S.}~\bibnamefont
  {{Nascimb{\`e}ne}}},\ }\bibfield  {title} {\bibinfo {title} {{Quantum
  simulations with ultracold quantum gases}},\ }\href
  {https://doi.org/10.1038/nphys2259} {\bibfield  {journal} {\bibinfo
  {journal} {Nat. Phys.}\ }\textbf {\bibinfo {volume} {8}},\ \bibinfo {pages}
  {267} (\bibinfo {year} {2012})}\BibitemShut {NoStop}%
\bibitem [{\citenamefont {{Haller}}\ \emph {et~al.}(2015)\citenamefont
  {{Haller}}, \citenamefont {{Hudson}}, \citenamefont {{Kelly}}, \citenamefont
  {{Cotta}}, \citenamefont {{Peaudecerf}}, \citenamefont {{Bruce}},\ and\
  \citenamefont {{Kuhr}}}]{Haller2015}%
  \BibitemOpen
  \bibfield  {author} {\bibinfo {author} {\bibfnamefont {E.}~\bibnamefont
  {{Haller}}}, \bibinfo {author} {\bibfnamefont {J.}~\bibnamefont {{Hudson}}},
  \bibinfo {author} {\bibfnamefont {A.}~\bibnamefont {{Kelly}}}, \bibinfo
  {author} {\bibfnamefont {D.~A.}\ \bibnamefont {{Cotta}}}, \bibinfo {author}
  {\bibfnamefont {B.}~\bibnamefont {{Peaudecerf}}}, \bibinfo {author}
  {\bibfnamefont {G.~D.}\ \bibnamefont {{Bruce}}},\ and\ \bibinfo {author}
  {\bibfnamefont {S.}~\bibnamefont {{Kuhr}}},\ }\bibfield  {title} {\bibinfo
  {title} {{Single-atom imaging of fermions in a quantum-gas microscope}},\
  }\href {https://doi.org/10.1038/nphys3403} {\bibfield  {journal} {\bibinfo
  {journal} {Nat. Phys.}\ }\textbf {\bibinfo {volume} {11}},\ \bibinfo {pages}
  {738} (\bibinfo {year} {2015})}\BibitemShut {NoStop}%
\bibitem [{\citenamefont {{Islam}}\ \emph {et~al.}(2015)\citenamefont
  {{Islam}}, \citenamefont {{Ma}}, \citenamefont {{Preiss}}, \citenamefont
  {{Eric Tai}}, \citenamefont {{Lukin}}, \citenamefont {{Rispoli}},\ and\
  \citenamefont {{Greiner}}}]{Islam2015}%
  \BibitemOpen
  \bibfield  {author} {\bibinfo {author} {\bibfnamefont {R.}~\bibnamefont
  {{Islam}}}, \bibinfo {author} {\bibfnamefont {R.}~\bibnamefont {{Ma}}},
  \bibinfo {author} {\bibfnamefont {P.~M.}\ \bibnamefont {{Preiss}}}, \bibinfo
  {author} {\bibfnamefont {M.}~\bibnamefont {{Eric Tai}}}, \bibinfo {author}
  {\bibfnamefont {A.}~\bibnamefont {{Lukin}}}, \bibinfo {author} {\bibfnamefont
  {M.}~\bibnamefont {{Rispoli}}},\ and\ \bibinfo {author} {\bibfnamefont
  {M.}~\bibnamefont {{Greiner}}},\ }\bibfield  {title} {\bibinfo {title}
  {{Measuring entanglement entropy in a quantum many-body system}},\ }\href
  {https://doi.org/10.1038/nature15750} {\bibfield  {journal} {\bibinfo
  {journal} {Nature}\ }\textbf {\bibinfo {volume} {528}},\ \bibinfo {pages}
  {77} (\bibinfo {year} {2015})}\BibitemShut {NoStop}%
\bibitem [{\citenamefont {{Parsons}}\ \emph {et~al.}(2016)\citenamefont
  {{Parsons}}, \citenamefont {{Mazurenko}}, \citenamefont {{Chiu}},
  \citenamefont {{Ji}}, \citenamefont {{Greif}},\ and\ \citenamefont
  {{Greiner}}}]{Parsons_2016}%
  \BibitemOpen
  \bibfield  {author} {\bibinfo {author} {\bibfnamefont {M.~F.}\ \bibnamefont
  {{Parsons}}}, \bibinfo {author} {\bibfnamefont {A.}~\bibnamefont
  {{Mazurenko}}}, \bibinfo {author} {\bibfnamefont {C.~S.}\ \bibnamefont
  {{Chiu}}}, \bibinfo {author} {\bibfnamefont {G.}~\bibnamefont {{Ji}}},
  \bibinfo {author} {\bibfnamefont {D.}~\bibnamefont {{Greif}}},\ and\ \bibinfo
  {author} {\bibfnamefont {M.}~\bibnamefont {{Greiner}}},\ }\bibfield  {title}
  {\bibinfo {title} {{Site-resolved measurement of the spin-correlation
  function in the Fermi-Hubbard model}},\ }\href
  {https://doi.org/10.1126/science.aag1430} {\bibfield  {journal} {\bibinfo
  {journal} {Science}\ }\textbf {\bibinfo {volume} {353}},\ \bibinfo {pages}
  {1253} (\bibinfo {year} {2016})}\BibitemShut {NoStop}%
\bibitem [{\citenamefont {{Choi}}\ \emph {et~al.}(2016)\citenamefont {{Choi}},
  \citenamefont {{Hild}}, \citenamefont {{Zeiher}}, \citenamefont
  {{Schau{\ss}}}, \citenamefont {{Rubio-Abadal}}, \citenamefont {{Yefsah}},
  \citenamefont {{Khemani}}, \citenamefont {{Huse}}, \citenamefont {{Bloch}},\
  and\ \citenamefont {{Gross}}}]{Choi2016}%
  \BibitemOpen
  \bibfield  {author} {\bibinfo {author} {\bibfnamefont {J.-y.}\ \bibnamefont
  {{Choi}}}, \bibinfo {author} {\bibfnamefont {S.}~\bibnamefont {{Hild}}},
  \bibinfo {author} {\bibfnamefont {J.}~\bibnamefont {{Zeiher}}}, \bibinfo
  {author} {\bibfnamefont {P.}~\bibnamefont {{Schau{\ss}}}}, \bibinfo {author}
  {\bibfnamefont {A.}~\bibnamefont {{Rubio-Abadal}}}, \bibinfo {author}
  {\bibfnamefont {T.}~\bibnamefont {{Yefsah}}}, \bibinfo {author}
  {\bibfnamefont {V.}~\bibnamefont {{Khemani}}}, \bibinfo {author}
  {\bibfnamefont {D.~A.}\ \bibnamefont {{Huse}}}, \bibinfo {author}
  {\bibfnamefont {I.}~\bibnamefont {{Bloch}}},\ and\ \bibinfo {author}
  {\bibfnamefont {C.}~\bibnamefont {{Gross}}},\ }\bibfield  {title} {\bibinfo
  {title} {{Exploring the many-body localization transition in two
  dimensions}},\ }\href {https://doi.org/10.1126/science.aaf8834} {\bibfield
  {journal} {\bibinfo  {journal} {Science}\ }\textbf {\bibinfo {volume}
  {352}},\ \bibinfo {pages} {1547} (\bibinfo {year} {2016})}\BibitemShut
  {NoStop}%
\bibitem [{\citenamefont {{Mazurenko}}\ \emph {et~al.}(2017)\citenamefont
  {{Mazurenko}}, \citenamefont {{Chiu}}, \citenamefont {{Ji}}, \citenamefont
  {{Parsons}}, \citenamefont {{Kan{\'a}sz-Nagy}}, \citenamefont {{Schmidt}},
  \citenamefont {{Grusdt}}, \citenamefont {{Demler}}, \citenamefont {{Greif}},\
  and\ \citenamefont {{Greiner}}}]{Mazurenko2017}%
  \BibitemOpen
  \bibfield  {author} {\bibinfo {author} {\bibfnamefont {A.}~\bibnamefont
  {{Mazurenko}}}, \bibinfo {author} {\bibfnamefont {C.~S.}\ \bibnamefont
  {{Chiu}}}, \bibinfo {author} {\bibfnamefont {G.}~\bibnamefont {{Ji}}},
  \bibinfo {author} {\bibfnamefont {M.~F.}\ \bibnamefont {{Parsons}}}, \bibinfo
  {author} {\bibfnamefont {M.}~\bibnamefont {{Kan{\'a}sz-Nagy}}}, \bibinfo
  {author} {\bibfnamefont {R.}~\bibnamefont {{Schmidt}}}, \bibinfo {author}
  {\bibfnamefont {F.}~\bibnamefont {{Grusdt}}}, \bibinfo {author}
  {\bibfnamefont {E.}~\bibnamefont {{Demler}}}, \bibinfo {author}
  {\bibfnamefont {D.}~\bibnamefont {{Greif}}},\ and\ \bibinfo {author}
  {\bibfnamefont {M.}~\bibnamefont {{Greiner}}},\ }\bibfield  {title} {\bibinfo
  {title} {{A cold-atom Fermi-Hubbard antiferromagnet}},\ }\href
  {https://doi.org/10.1038/nature22362} {\bibfield  {journal} {\bibinfo
  {journal} {Nature}\ }\textbf {\bibinfo {volume} {545}},\ \bibinfo {pages}
  {462} (\bibinfo {year} {2017})}\BibitemShut {NoStop}%
\bibitem [{\citenamefont {{Gritsev}}\ \emph {et~al.}(2006)\citenamefont
  {{Gritsev}}, \citenamefont {{Altman}}, \citenamefont {{Demler}},\ and\
  \citenamefont {{Polkovnikov}}}]{Gritsev_2006}%
  \BibitemOpen
  \bibfield  {author} {\bibinfo {author} {\bibfnamefont {V.}~\bibnamefont
  {{Gritsev}}}, \bibinfo {author} {\bibfnamefont {E.}~\bibnamefont {{Altman}}},
  \bibinfo {author} {\bibfnamefont {E.}~\bibnamefont {{Demler}}},\ and\
  \bibinfo {author} {\bibfnamefont {A.}~\bibnamefont {{Polkovnikov}}},\
  }\bibfield  {title} {\bibinfo {title} {{Full quantum distribution of contrast
  in interference experiments between interacting one-dimensional Bose
  liquids}},\ }\href {https://doi.org/10.1038/nphys410} {\bibfield  {journal}
  {\bibinfo  {journal} {Nat. Phys.}\ }\textbf {\bibinfo {volume} {2}},\
  \bibinfo {pages} {705} (\bibinfo {year} {2006})}\BibitemShut {NoStop}%
\bibitem [{\citenamefont {{Hofferberth}}\ \emph {et~al.}(2008)\citenamefont
  {{Hofferberth}}, \citenamefont {{Lesanovsky}}, \citenamefont {{Schumm}},
  \citenamefont {{Imambekov}}, \citenamefont {{Gritsev}}, \citenamefont
  {{Demler}},\ and\ \citenamefont {{Schmiedmayer}}}]{Hofferberth_2008}%
  \BibitemOpen
  \bibfield  {author} {\bibinfo {author} {\bibfnamefont {S.}~\bibnamefont
  {{Hofferberth}}}, \bibinfo {author} {\bibfnamefont {I.}~\bibnamefont
  {{Lesanovsky}}}, \bibinfo {author} {\bibfnamefont {T.}~\bibnamefont
  {{Schumm}}}, \bibinfo {author} {\bibfnamefont {A.}~\bibnamefont
  {{Imambekov}}}, \bibinfo {author} {\bibfnamefont {V.}~\bibnamefont
  {{Gritsev}}}, \bibinfo {author} {\bibfnamefont {E.}~\bibnamefont
  {{Demler}}},\ and\ \bibinfo {author} {\bibfnamefont {J.}~\bibnamefont
  {{Schmiedmayer}}},\ }\bibfield  {title} {\bibinfo {title} {{Probing quantum
  and thermal noise in an interacting many-body system}},\ }\href
  {https://doi.org/10.1038/nphys941} {\bibfield  {journal} {\bibinfo  {journal}
  {Nat. Phys.}\ }\textbf {\bibinfo {volume} {4}},\ \bibinfo {pages} {489}
  (\bibinfo {year} {2008})}\BibitemShut {NoStop}%
\bibitem [{\citenamefont {{Ashida}}\ and\ \citenamefont
  {{Ueda}}(2017)}]{Ashida_2018}%
  \BibitemOpen
  \bibfield  {author} {\bibinfo {author} {\bibfnamefont {Y.}~\bibnamefont
  {{Ashida}}}\ and\ \bibinfo {author} {\bibfnamefont {M.}~\bibnamefont
  {{Ueda}}},\ }\bibfield  {title} {\bibinfo {title} {{Full-Counting
  Many-Particle Dynamics: Nonlocal and Chiral Propagation of Correlations}},\
  }\href {https://doi.org/10.48550/arXiv.1709.03704} {\bibfield  {journal}
  {\bibinfo  {journal} {arXiv e-prints}\ ,\ \bibinfo {eid} {arXiv:1709.03704}}
  (\bibinfo {year} {2017})}\BibitemShut {NoStop}%
\bibitem [{\citenamefont {{Ridley}}\ \emph {et~al.}(2018)\citenamefont
  {{Ridley}}, \citenamefont {{Singh}}, \citenamefont {{Gull}},\ and\
  \citenamefont {{Cohen}}}]{Ridley2018}%
  \BibitemOpen
  \bibfield  {author} {\bibinfo {author} {\bibfnamefont {M.}~\bibnamefont
  {{Ridley}}}, \bibinfo {author} {\bibfnamefont {V.~N.}\ \bibnamefont
  {{Singh}}}, \bibinfo {author} {\bibfnamefont {E.}~\bibnamefont {{Gull}}},\
  and\ \bibinfo {author} {\bibfnamefont {G.}~\bibnamefont {{Cohen}}},\
  }\bibfield  {title} {\bibinfo {title} {{Numerically exact full counting
  statistics of the nonequilibrium Anderson impurity model}},\ }\href
  {https://doi.org/10.1103/PhysRevB.97.115109} {\bibfield  {journal} {\bibinfo
  {journal} {\prb}\ }\textbf {\bibinfo {volume} {97}},\ \bibinfo {eid} {115109}
  (\bibinfo {year} {2018})}\BibitemShut {NoStop}%
\bibitem [{\citenamefont {{Ridley}}\ \emph {et~al.}(2019)\citenamefont
  {{Ridley}}, \citenamefont {{Galperin}}, \citenamefont {{Gull}},\ and\
  \citenamefont {{Cohen}}}]{Ridley2019}%
  \BibitemOpen
  \bibfield  {author} {\bibinfo {author} {\bibfnamefont {M.}~\bibnamefont
  {{Ridley}}}, \bibinfo {author} {\bibfnamefont {M.}~\bibnamefont
  {{Galperin}}}, \bibinfo {author} {\bibfnamefont {E.}~\bibnamefont {{Gull}}},\
  and\ \bibinfo {author} {\bibfnamefont {G.}~\bibnamefont {{Cohen}}},\
  }\bibfield  {title} {\bibinfo {title} {{Numerically exact full counting
  statistics of the energy current in the Kondo regime}},\ }\href
  {https://doi.org/10.1103/PhysRevB.100.165127} {\bibfield  {journal} {\bibinfo
   {journal} {\prb}\ }\textbf {\bibinfo {volume} {100}},\ \bibinfo {eid}
  {165127} (\bibinfo {year} {2019})}\BibitemShut {NoStop}%
\bibitem [{\citenamefont {{Bastianello}}\ and\ \citenamefont
  {{Piroli}}(2018)}]{Bastianello_2018}%
  \BibitemOpen
  \bibfield  {author} {\bibinfo {author} {\bibfnamefont {A.}~\bibnamefont
  {{Bastianello}}}\ and\ \bibinfo {author} {\bibfnamefont {L.}~\bibnamefont
  {{Piroli}}},\ }\bibfield  {title} {\bibinfo {title} {{From the sinh-Gordon
  field theory to the one-dimensional Bose gas: exact local correlations and
  full counting statistics}},\ }\href
  {https://doi.org/10.1088/1742-5468/aaeb48} {\bibfield  {journal} {\bibinfo
  {journal} {J. Stat. Mech.}\ }\textbf {\bibinfo {volume} {11}},\ \bibinfo
  {pages} {113104} (\bibinfo {year} {2018})}\BibitemShut {NoStop}%
\bibitem [{\citenamefont {{McCulloch}}\ \emph {et~al.}(2023)\citenamefont
  {{McCulloch}}, \citenamefont {{De Nardis}}, \citenamefont
  {{Gopalakrishnan}},\ and\ \citenamefont {{Vasseur}}}]{McCulloch2023}%
  \BibitemOpen
  \bibfield  {author} {\bibinfo {author} {\bibfnamefont {E.}~\bibnamefont
  {{McCulloch}}}, \bibinfo {author} {\bibfnamefont {J.}~\bibnamefont {{De
  Nardis}}}, \bibinfo {author} {\bibfnamefont {S.}~\bibnamefont
  {{Gopalakrishnan}}},\ and\ \bibinfo {author} {\bibfnamefont {R.}~\bibnamefont
  {{Vasseur}}},\ }\bibfield  {title} {\bibinfo {title} {{Full Counting
  Statistics of Charge in Chaotic Many-body Quantum Systems}},\ }\href
  {https://doi.org/10.48550/arXiv.2302.01355} {\bibfield  {journal} {\bibinfo
  {journal} {arXiv e-prints}\ ,\ \bibinfo {eid} {arXiv:2302.01355}} (\bibinfo
  {year} {2023})}\BibitemShut {NoStop}%
\bibitem [{\citenamefont {{Bertini}}\ \emph {et~al.}(2022)\citenamefont
  {{Bertini}}, \citenamefont {{Calabrese}}, \citenamefont {{Collura}},
  \citenamefont {{Klobas}},\ and\ \citenamefont {{Rylands}}}]{Bertini2023}%
  \BibitemOpen
  \bibfield  {author} {\bibinfo {author} {\bibfnamefont {B.}~\bibnamefont
  {{Bertini}}}, \bibinfo {author} {\bibfnamefont {P.}~\bibnamefont
  {{Calabrese}}}, \bibinfo {author} {\bibfnamefont {M.}~\bibnamefont
  {{Collura}}}, \bibinfo {author} {\bibfnamefont {K.}~\bibnamefont
  {{Klobas}}},\ and\ \bibinfo {author} {\bibfnamefont {C.}~\bibnamefont
  {{Rylands}}},\ }\bibfield  {title} {\bibinfo {title} {{Nonequilibrium Full
  Counting Statistics and Symmetry-Resolved Entanglement from Space-Time
  Duality}},\ }\href {https://doi.org/10.48550/arXiv.2212.06188} {\bibfield
  {journal} {\bibinfo  {journal} {arXiv e-prints}\ ,\ \bibinfo {eid}
  {arXiv:2212.06188}} (\bibinfo {year} {2022})}\BibitemShut {NoStop}%
\bibitem [{\citenamefont {{Gopalakrishnan}}\ \emph {et~al.}(2022)\citenamefont
  {{Gopalakrishnan}}, \citenamefont {{Morningstar}}, \citenamefont
  {{Vasseur}},\ and\ \citenamefont {{Khemani}}}]{XXZFCS}%
  \BibitemOpen
  \bibfield  {author} {\bibinfo {author} {\bibfnamefont {S.}~\bibnamefont
  {{Gopalakrishnan}}}, \bibinfo {author} {\bibfnamefont {A.}~\bibnamefont
  {{Morningstar}}}, \bibinfo {author} {\bibfnamefont {R.}~\bibnamefont
  {{Vasseur}}},\ and\ \bibinfo {author} {\bibfnamefont {V.}~\bibnamefont
  {{Khemani}}},\ }\bibfield  {title} {\bibinfo {title} {{Distinct universality
  classes of diffusive transport from full counting statistics}},\ }\href
  {https://doi.org/10.48550/arXiv.2203.09526} {\bibfield  {journal} {\bibinfo
  {journal} {arXiv e-prints}\ ,\ \bibinfo {eid} {arXiv:2203.09526}} (\bibinfo
  {year} {2022})}\BibitemShut {NoStop}%
\bibitem [{\citenamefont {{Wei}}\ \emph {et~al.}(2022)\citenamefont {{Wei}},
  \citenamefont {{Rubio-Abadal}}, \citenamefont {{Ye}}, \citenamefont
  {{Machado}}, \citenamefont {{Kemp}}, \citenamefont {{Srakaew}}, \citenamefont
  {{Hollerith}}, \citenamefont {{Rui}}, \citenamefont {{Gopalakrishnan}},
  \citenamefont {{Yao}}, \citenamefont {{Bloch}},\ and\ \citenamefont
  {{Zeiher}}}]{Wei2022}%
  \BibitemOpen
  \bibfield  {author} {\bibinfo {author} {\bibfnamefont {D.}~\bibnamefont
  {{Wei}}}, \bibinfo {author} {\bibfnamefont {A.}~\bibnamefont
  {{Rubio-Abadal}}}, \bibinfo {author} {\bibfnamefont {B.}~\bibnamefont
  {{Ye}}}, \bibinfo {author} {\bibfnamefont {F.}~\bibnamefont {{Machado}}},
  \bibinfo {author} {\bibfnamefont {J.}~\bibnamefont {{Kemp}}}, \bibinfo
  {author} {\bibfnamefont {K.}~\bibnamefont {{Srakaew}}}, \bibinfo {author}
  {\bibfnamefont {S.}~\bibnamefont {{Hollerith}}}, \bibinfo {author}
  {\bibfnamefont {J.}~\bibnamefont {{Rui}}}, \bibinfo {author} {\bibfnamefont
  {S.}~\bibnamefont {{Gopalakrishnan}}}, \bibinfo {author} {\bibfnamefont
  {N.~Y.}\ \bibnamefont {{Yao}}}, \bibinfo {author} {\bibfnamefont
  {I.}~\bibnamefont {{Bloch}}},\ and\ \bibinfo {author} {\bibfnamefont
  {J.}~\bibnamefont {{Zeiher}}},\ }\bibfield  {title} {\bibinfo {title}
  {{Quantum gas microscopy of Kardar-Parisi-Zhang superdiffusion}},\ }\href
  {https://doi.org/10.1126/science.abk2397} {\bibfield  {journal} {\bibinfo
  {journal} {Science}\ }\textbf {\bibinfo {volume} {376}},\ \bibinfo {pages}
  {716} (\bibinfo {year} {2022})}\BibitemShut {NoStop}%
\bibitem [{\citenamefont {{De Nardis}}\ \emph {et~al.}(2022)\citenamefont {{De
  Nardis}}, \citenamefont {{Gopalakrishnan}},\ and\ \citenamefont
  {{Vasseur}}}]{DeNardis2022}%
  \BibitemOpen
  \bibfield  {author} {\bibinfo {author} {\bibfnamefont {J.}~\bibnamefont {{De
  Nardis}}}, \bibinfo {author} {\bibfnamefont {S.}~\bibnamefont
  {{Gopalakrishnan}}},\ and\ \bibinfo {author} {\bibfnamefont {R.}~\bibnamefont
  {{Vasseur}}},\ }\bibfield  {title} {\bibinfo {title} {{Non-linear fluctuating
  hydrodynamics for KPZ scaling in isotropic spin chains}},\ }\href
  {https://doi.org/10.48550/arXiv.2212.03696} {\bibfield  {journal} {\bibinfo
  {journal} {arXiv e-prints}\ ,\ \bibinfo {eid} {arXiv:2212.03696}} (\bibinfo
  {year} {2022})}\BibitemShut {NoStop}%
\bibitem [{\citenamefont {{Krajnik}}\ \emph
  {et~al.}(2022{\natexlab{a}})\citenamefont {{Krajnik}}, \citenamefont
  {{Schmidt}}, \citenamefont {{Pasquier}}, \citenamefont {{Ilievski}},\ and\
  \citenamefont {{Prosen}}}]{Krajnik2022a}%
  \BibitemOpen
  \bibfield  {author} {\bibinfo {author} {\bibfnamefont {{\v{Z}}.}~\bibnamefont
  {{Krajnik}}}, \bibinfo {author} {\bibfnamefont {J.}~\bibnamefont
  {{Schmidt}}}, \bibinfo {author} {\bibfnamefont {V.}~\bibnamefont
  {{Pasquier}}}, \bibinfo {author} {\bibfnamefont {E.}~\bibnamefont
  {{Ilievski}}},\ and\ \bibinfo {author} {\bibfnamefont {T.}~\bibnamefont
  {{Prosen}}},\ }\bibfield  {title} {\bibinfo {title} {{Exact Anomalous Current
  Fluctuations in a Deterministic Interacting Model}},\ }\href
  {https://doi.org/10.1103/PhysRevLett.128.160601} {\bibfield  {journal}
  {\bibinfo  {journal} {\prl}\ }\textbf {\bibinfo {volume} {128}},\ \bibinfo
  {eid} {160601} (\bibinfo {year} {2022}{\natexlab{a}})}\BibitemShut {NoStop}%
\bibitem [{\citenamefont {{Krajnik}}\ \emph
  {et~al.}(2022{\natexlab{b}})\citenamefont {{Krajnik}}, \citenamefont
  {{Ilievski}},\ and\ \citenamefont {{Prosen}}}]{Krajnik2022b}%
  \BibitemOpen
  \bibfield  {author} {\bibinfo {author} {\bibfnamefont {{\v{Z}}.}~\bibnamefont
  {{Krajnik}}}, \bibinfo {author} {\bibfnamefont {E.}~\bibnamefont
  {{Ilievski}}},\ and\ \bibinfo {author} {\bibfnamefont {T.}~\bibnamefont
  {{Prosen}}},\ }\bibfield  {title} {\bibinfo {title} {{Absence of Normal
  Fluctuations in an Integrable Magnet}},\ }\href
  {https://doi.org/10.1103/PhysRevLett.128.090604} {\bibfield  {journal}
  {\bibinfo  {journal} {\prl}\ }\textbf {\bibinfo {volume} {128}},\ \bibinfo
  {eid} {090604} (\bibinfo {year} {2022}{\natexlab{b}})}\BibitemShut {NoStop}%
\bibitem [{\citenamefont {{Krajnik}}\ \emph
  {et~al.}(2022{\natexlab{c}})\citenamefont {{Krajnik}}, \citenamefont
  {{Schmidt}}, \citenamefont {{Pasquier}}, \citenamefont {{Prosen}},\ and\
  \citenamefont {{Ilievski}}}]{Krajnik2022c}%
  \BibitemOpen
  \bibfield  {author} {\bibinfo {author} {\bibfnamefont {{\v{Z}}.}~\bibnamefont
  {{Krajnik}}}, \bibinfo {author} {\bibfnamefont {J.}~\bibnamefont
  {{Schmidt}}}, \bibinfo {author} {\bibfnamefont {V.}~\bibnamefont
  {{Pasquier}}}, \bibinfo {author} {\bibfnamefont {T.}~\bibnamefont
  {{Prosen}}},\ and\ \bibinfo {author} {\bibfnamefont {E.}~\bibnamefont
  {{Ilievski}}},\ }\bibfield  {title} {\bibinfo {title} {{Universal anomalous
  fluctuations in charged single-file systems}},\ }\href
  {https://doi.org/10.48550/arXiv.2208.01463} {\bibfield  {journal} {\bibinfo
  {journal} {arXiv e-prints}\ ,\ \bibinfo {eid} {arXiv:2208.01463}} (\bibinfo
  {year} {2022}{\natexlab{c}})}\BibitemShut {NoStop}%
\bibitem [{\citenamefont {Ba\~nuls}\ and\ \citenamefont
  {Garrahan}(2019)}]{PhysRevLett.123.200601}%
  \BibitemOpen
  \bibfield  {author} {\bibinfo {author} {\bibfnamefont {M.~C.}\ \bibnamefont
  {Ba\~nuls}}\ and\ \bibinfo {author} {\bibfnamefont {J.~P.}\ \bibnamefont
  {Garrahan}},\ }\bibfield  {title} {\bibinfo {title} {{Using Matrix Product
  States to Study the Dynamical Large Deviations of Kinetically Constrained
  Models}},\ }\href {https://doi.org/10.1103/PhysRevLett.123.200601} {\bibfield
   {journal} {\bibinfo  {journal} {Phys. Rev. Lett.}\ }\textbf {\bibinfo
  {volume} {123}},\ \bibinfo {pages} {200601} (\bibinfo {year}
  {2019})}\BibitemShut {NoStop}%
\bibitem [{\citenamefont {Causer}\ \emph {et~al.}(2022)\citenamefont {Causer},
  \citenamefont {Ba\~nuls},\ and\ \citenamefont
  {Garrahan}}]{PhysRevLett.128.090605}%
  \BibitemOpen
  \bibfield  {author} {\bibinfo {author} {\bibfnamefont {L.}~\bibnamefont
  {Causer}}, \bibinfo {author} {\bibfnamefont {M.~C.}\ \bibnamefont
  {Ba\~nuls}},\ and\ \bibinfo {author} {\bibfnamefont {J.~P.}\ \bibnamefont
  {Garrahan}},\ }\bibfield  {title} {\bibinfo {title} {{Finite Time Large
  Deviations via Matrix Product States}},\ }\href
  {https://doi.org/10.1103/PhysRevLett.128.090605} {\bibfield  {journal}
  {\bibinfo  {journal} {Phys. Rev. Lett.}\ }\textbf {\bibinfo {volume} {128}},\
  \bibinfo {pages} {090605} (\bibinfo {year} {2022})}\BibitemShut {NoStop}%
\bibitem [{\citenamefont {Fowler}\ \emph {et~al.}(2012)\citenamefont {Fowler},
  \citenamefont {Mariantoni}, \citenamefont {Martinis},\ and\ \citenamefont
  {Cleland}}]{PhysRevA.86.032324}%
  \BibitemOpen
  \bibfield  {author} {\bibinfo {author} {\bibfnamefont {A.~G.}\ \bibnamefont
  {Fowler}}, \bibinfo {author} {\bibfnamefont {M.}~\bibnamefont {Mariantoni}},
  \bibinfo {author} {\bibfnamefont {J.~M.}\ \bibnamefont {Martinis}},\ and\
  \bibinfo {author} {\bibfnamefont {A.~N.}\ \bibnamefont {Cleland}},\
  }\bibfield  {title} {\bibinfo {title} {Surface codes: Towards practical
  large-scale quantum computation},\ }\href
  {https://doi.org/10.1103/PhysRevA.86.032324} {\bibfield  {journal} {\bibinfo
  {journal} {Phys. Rev. A}\ }\textbf {\bibinfo {volume} {86}},\ \bibinfo
  {pages} {032324} (\bibinfo {year} {2012})}\BibitemShut {NoStop}%
\bibitem [{\citenamefont {Xu}\ and\ \citenamefont {del
  Campo}(2019)}]{xu2019probing}%
  \BibitemOpen
  \bibfield  {author} {\bibinfo {author} {\bibfnamefont {Z.}~\bibnamefont
  {Xu}}\ and\ \bibinfo {author} {\bibfnamefont {A.}~\bibnamefont {del Campo}},\
  }\bibfield  {title} {\bibinfo {title} {{Probing the Full Distribution of
  Many-Body Observables By Single-Qubit Interferometry}},\ }\href
  {https://doi.org/10.1103/PhysRevLett.122.160602} {\bibfield  {journal}
  {\bibinfo  {journal} {Phys. Rev. Lett.}\ }\textbf {\bibinfo {volume} {122}},\
  \bibinfo {pages} {160602} (\bibinfo {year} {2019})}\BibitemShut {NoStop}%
\bibitem [{\citenamefont {{White}}\ \emph {et~al.}(2018)\citenamefont
  {{White}}, \citenamefont {{Zaletel}}, \citenamefont {{Mong}},\ and\
  \citenamefont {{Refael}}}]{White2018}%
  \BibitemOpen
  \bibfield  {author} {\bibinfo {author} {\bibfnamefont {C.~D.}\ \bibnamefont
  {{White}}}, \bibinfo {author} {\bibfnamefont {M.}~\bibnamefont {{Zaletel}}},
  \bibinfo {author} {\bibfnamefont {R.~S.~K.}\ \bibnamefont {{Mong}}},\ and\
  \bibinfo {author} {\bibfnamefont {G.}~\bibnamefont {{Refael}}},\ }\bibfield
  {title} {\bibinfo {title} {{Quantum dynamics of thermalizing systems}},\
  }\href {https://doi.org/10.1103/PhysRevB.97.035127} {\bibfield  {journal}
  {\bibinfo  {journal} {\prb}\ }\textbf {\bibinfo {volume} {97}},\ \bibinfo
  {eid} {035127} (\bibinfo {year} {2018})}\BibitemShut {NoStop}%
\bibitem [{\citenamefont {Vanicat}\ \emph {et~al.}(2018)\citenamefont
  {Vanicat}, \citenamefont {Zadnik},\ and\ \citenamefont
  {Prosen}}]{vanicat2018integrable}%
  \BibitemOpen
  \bibfield  {author} {\bibinfo {author} {\bibfnamefont {M.}~\bibnamefont
  {Vanicat}}, \bibinfo {author} {\bibfnamefont {L.}~\bibnamefont {Zadnik}},\
  and\ \bibinfo {author} {\bibfnamefont {T.}~\bibnamefont {Prosen}},\
  }\bibfield  {title} {\bibinfo {title} {{Integrable Trotterization: Local
  Conservation Laws and Boundary Driving}},\ }\href
  {https://doi.org/10.1103/PhysRevLett.121.030606} {\bibfield  {journal}
  {\bibinfo  {journal} {Phys. Rev. Lett.}\ }\textbf {\bibinfo {volume} {121}},\
  \bibinfo {pages} {030606} (\bibinfo {year} {2018})}\BibitemShut {NoStop}%
\bibitem [{\citenamefont {Ljubotina}\ \emph
  {et~al.}(2019{\natexlab{a}})\citenamefont {Ljubotina}, \citenamefont
  {Zadnik},\ and\ \citenamefont {Prosen}}]{ljubotina2019ballistic}%
  \BibitemOpen
  \bibfield  {author} {\bibinfo {author} {\bibfnamefont {M.}~\bibnamefont
  {Ljubotina}}, \bibinfo {author} {\bibfnamefont {L.}~\bibnamefont {Zadnik}},\
  and\ \bibinfo {author} {\bibfnamefont {T.}~\bibnamefont {Prosen}},\
  }\bibfield  {title} {\bibinfo {title} {{Ballistic Spin Transport in a
  Periodically Driven Integrable Quantum System}},\ }\href
  {https://doi.org/10.1103/PhysRevLett.122.150605} {\bibfield  {journal}
  {\bibinfo  {journal} {Phys. Rev. Lett.}\ }\textbf {\bibinfo {volume} {122}},\
  \bibinfo {pages} {150605} (\bibinfo {year} {2019}{\natexlab{a}})}\BibitemShut
  {NoStop}%
\bibitem [{\citenamefont {{\v{Z}}nidari{\v{c}}}(2011)}]{znidarivc2011spin}%
  \BibitemOpen
  \bibfield  {author} {\bibinfo {author} {\bibfnamefont {M.}~\bibnamefont
  {{\v{Z}}nidari{\v{c}}}},\ }\bibfield  {title} {\bibinfo {title} {{Spin
  transport in a one-dimensional anisotropic Heisenberg model}},\ }\href
  {https://doi.org/10.1103/PhysRevLett.106.220601} {\bibfield  {journal}
  {\bibinfo  {journal} {Phys. Rev. Lett.}\ }\textbf {\bibinfo {volume} {106}},\
  \bibinfo {pages} {220601} (\bibinfo {year} {2011})}\BibitemShut {NoStop}%
\bibitem [{\citenamefont {Ljubotina}\ \emph {et~al.}(2017)\citenamefont
  {Ljubotina}, \citenamefont {{\v{Z}}nidari{\v{c}}},\ and\ \citenamefont
  {Prosen}}]{ljubotina2017spin}%
  \BibitemOpen
  \bibfield  {author} {\bibinfo {author} {\bibfnamefont {M.}~\bibnamefont
  {Ljubotina}}, \bibinfo {author} {\bibfnamefont {M.}~\bibnamefont
  {{\v{Z}}nidari{\v{c}}}},\ and\ \bibinfo {author} {\bibfnamefont
  {T.}~\bibnamefont {Prosen}},\ }\bibfield  {title} {\bibinfo {title} {Spin
  diffusion from an inhomogeneous quench in an integrable system},\ }\href
  {https://doi.org/10.1038/ncomms16117} {\bibfield  {journal} {\bibinfo
  {journal} {Nat. Commun.}\ }\textbf {\bibinfo {volume} {8}},\ \bibinfo {pages}
  {16117} (\bibinfo {year} {2017})}\BibitemShut {NoStop}%
\bibitem [{\citenamefont {Ljubotina}\ \emph
  {et~al.}(2019{\natexlab{b}})\citenamefont {Ljubotina}, \citenamefont
  {{\v{Z}}nidari{\v{c}}},\ and\ \citenamefont {Prosen}}]{ljubotina2019kardar}%
  \BibitemOpen
  \bibfield  {author} {\bibinfo {author} {\bibfnamefont {M.}~\bibnamefont
  {Ljubotina}}, \bibinfo {author} {\bibfnamefont {M.}~\bibnamefont
  {{\v{Z}}nidari{\v{c}}}},\ and\ \bibinfo {author} {\bibfnamefont
  {T.}~\bibnamefont {Prosen}},\ }\bibfield  {title} {\bibinfo {title}
  {Kardar-parisi-zhang physics in the quantum heisenberg magnet},\ }\href
  {https://doi.org/10.1103/PhysRevLett.122.210602} {\bibfield  {journal}
  {\bibinfo  {journal} {Phys. Rev. Lett.}\ }\textbf {\bibinfo {volume} {122}},\
  \bibinfo {pages} {210602} (\bibinfo {year} {2019}{\natexlab{b}})}\BibitemShut
  {NoStop}%
\bibitem [{\citenamefont {Gopalakrishnan}\ and\ \citenamefont
  {Vasseur}(2019)}]{gopalakrishnan2019kinetic}%
  \BibitemOpen
  \bibfield  {author} {\bibinfo {author} {\bibfnamefont {S.}~\bibnamefont
  {Gopalakrishnan}}\ and\ \bibinfo {author} {\bibfnamefont {R.}~\bibnamefont
  {Vasseur}},\ }\bibfield  {title} {\bibinfo {title} {{Kinetic Theory of Spin
  Diffusion and Superdiffusion in $XXZ$ Spin Chains}},\ }\href
  {https://doi.org/10.1103/PhysRevLett.122.127202} {\bibfield  {journal}
  {\bibinfo  {journal} {Phys. Rev. Lett.}\ }\textbf {\bibinfo {volume} {122}},\
  \bibinfo {pages} {127202} (\bibinfo {year} {2019})}\BibitemShut {NoStop}%
\bibitem [{\citenamefont {De~Nardis}\ \emph {et~al.}(2019)\citenamefont
  {De~Nardis}, \citenamefont {Medenjak}, \citenamefont {Karrasch},\ and\
  \citenamefont {Ilievski}}]{de2019anomalous}%
  \BibitemOpen
  \bibfield  {author} {\bibinfo {author} {\bibfnamefont {J.}~\bibnamefont
  {De~Nardis}}, \bibinfo {author} {\bibfnamefont {M.}~\bibnamefont {Medenjak}},
  \bibinfo {author} {\bibfnamefont {C.}~\bibnamefont {Karrasch}},\ and\
  \bibinfo {author} {\bibfnamefont {E.}~\bibnamefont {Ilievski}},\ }\bibfield
  {title} {\bibinfo {title} {{Anomalous Spin Diffusion in One-Dimensional
  Antiferromagnets}},\ }\href {https://doi.org/10.1103/PhysRevLett.123.186601}
  {\bibfield  {journal} {\bibinfo  {journal} {Phys. Rev. Lett.}\ }\textbf
  {\bibinfo {volume} {123}},\ \bibinfo {pages} {186601} (\bibinfo {year}
  {2019})}\BibitemShut {NoStop}%
\bibitem [{\citenamefont {Gopalakrishnan}\ \emph {et~al.}(2019)\citenamefont
  {Gopalakrishnan}, \citenamefont {Vasseur},\ and\ \citenamefont
  {Ware}}]{gopalakrishnan2019anomalous}%
  \BibitemOpen
  \bibfield  {author} {\bibinfo {author} {\bibfnamefont {S.}~\bibnamefont
  {Gopalakrishnan}}, \bibinfo {author} {\bibfnamefont {R.}~\bibnamefont
  {Vasseur}},\ and\ \bibinfo {author} {\bibfnamefont {B.}~\bibnamefont
  {Ware}},\ }\bibfield  {title} {\bibinfo {title} {{Anomalous relaxation and
  the high-temperature structure factor of XXZ spin chains}},\ }\href
  {https://doi.org/10.1073/pnas.1906914116} {\bibfield  {journal} {\bibinfo
  {journal} {Proc. Natl. Acad. Sci. U.S.A.}\ }\textbf {\bibinfo {volume}
  {116}},\ \bibinfo {pages} {16250} (\bibinfo {year} {2019})}\BibitemShut
  {NoStop}%
\bibitem [{\citenamefont {Bulchandani}(2020)}]{bulchandani2020kardar}%
  \BibitemOpen
  \bibfield  {author} {\bibinfo {author} {\bibfnamefont {V.~B.}\ \bibnamefont
  {Bulchandani}},\ }\bibfield  {title} {\bibinfo {title} {{Kardar-Parisi-Zhang
  universality from soft gauge modes}},\ }\href
  {https://doi.org/10.1103/PhysRevB.101.041411} {\bibfield  {journal} {\bibinfo
   {journal} {Phys. Rev. B}\ }\textbf {\bibinfo {volume} {101}},\ \bibinfo
  {pages} {041411} (\bibinfo {year} {2020})}\BibitemShut {NoStop}%
\bibitem [{\citenamefont {Gopalakrishnan}\ and\ \citenamefont
  {Vasseur}(2023)}]{gopalakrishnan2022anomalous}%
  \BibitemOpen
  \bibfield  {author} {\bibinfo {author} {\bibfnamefont {S.}~\bibnamefont
  {Gopalakrishnan}}\ and\ \bibinfo {author} {\bibfnamefont {R.}~\bibnamefont
  {Vasseur}},\ }\bibfield  {title} {\bibinfo {title} {Anomalous transport from
  hot quasiparticles in interacting spin chains},\ }\href
  {https://doi.org/10.1088/1361-6633/acb36e} {\bibfield  {journal} {\bibinfo
  {journal} {Rep. Prog. Phys.}\ }\textbf {\bibinfo {volume} {86}},\ \bibinfo
  {pages} {036502} (\bibinfo {year} {2023})}\BibitemShut {NoStop}%
\bibitem [{\citenamefont {De~Nardis}\ \emph {et~al.}(2021)\citenamefont
  {De~Nardis}, \citenamefont {Gopalakrishnan}, \citenamefont {Vasseur},\ and\
  \citenamefont {Ware}}]{de2021stability}%
  \BibitemOpen
  \bibfield  {author} {\bibinfo {author} {\bibfnamefont {J.}~\bibnamefont
  {De~Nardis}}, \bibinfo {author} {\bibfnamefont {S.}~\bibnamefont
  {Gopalakrishnan}}, \bibinfo {author} {\bibfnamefont {R.}~\bibnamefont
  {Vasseur}},\ and\ \bibinfo {author} {\bibfnamefont {B.}~\bibnamefont
  {Ware}},\ }\bibfield  {title} {\bibinfo {title} {{Stability of Superdiffusion
  in Nearly Integrable Spin Chains}},\ }\href
  {https://doi.org/10.1103/PhysRevLett.127.057201} {\bibfield  {journal}
  {\bibinfo  {journal} {Phys. Rev. Lett.}\ }\textbf {\bibinfo {volume} {127}},\
  \bibinfo {pages} {057201} (\bibinfo {year} {2021})}\BibitemShut {NoStop}%
\bibitem [{\citenamefont {\relax Google Quantum~AI}\ and\ \citenamefont
  {Collaborators}(2023)}]{google2023}%
  \BibitemOpen
  \bibfield  {author} {\bibinfo {author} {\bibnamefont {\relax Google
  Quantum~AI}}\ and\ \bibinfo {author} {\bibnamefont {Collaborators}},\
  }\href@noop {} {\bibinfo {title} {{Dynamics of magnetization at infinite
  temperature in a Heisenberg spin chain}}} (\bibinfo {year}
  {2023})\BibitemShut {NoStop}%
\bibitem [{\citenamefont {T{\'o}th}\ and\ \citenamefont
  {Apellaniz}(2014)}]{toth2014quantum}%
  \BibitemOpen
  \bibfield  {author} {\bibinfo {author} {\bibfnamefont {G.}~\bibnamefont
  {T{\'o}th}}\ and\ \bibinfo {author} {\bibfnamefont {I.}~\bibnamefont
  {Apellaniz}},\ }\bibfield  {title} {\bibinfo {title} {Quantum metrology from
  a quantum information science perspective},\ }\href
  {https://doi.org/10.1088/1751-8113/47/42/424006} {\bibfield  {journal}
  {\bibinfo  {journal} {J. Phys. A: Math. Theor.}\ }\textbf {\bibinfo {volume}
  {47}},\ \bibinfo {pages} {424006} (\bibinfo {year} {2014})}\BibitemShut
  {NoStop}%
\bibitem [{\citenamefont {Kessler}\ \emph {et~al.}(2014)\citenamefont
  {Kessler}, \citenamefont {Lovchinsky}, \citenamefont {Sushkov},\ and\
  \citenamefont {Lukin}}]{PhysRevLett.112.150802}%
  \BibitemOpen
  \bibfield  {author} {\bibinfo {author} {\bibfnamefont {E.~M.}\ \bibnamefont
  {Kessler}}, \bibinfo {author} {\bibfnamefont {I.}~\bibnamefont {Lovchinsky}},
  \bibinfo {author} {\bibfnamefont {A.~O.}\ \bibnamefont {Sushkov}},\ and\
  \bibinfo {author} {\bibfnamefont {M.~D.}\ \bibnamefont {Lukin}},\ }\bibfield
  {title} {\bibinfo {title} {Quantum error correction for metrology},\ }\href
  {https://doi.org/10.1103/PhysRevLett.112.150802} {\bibfield  {journal}
  {\bibinfo  {journal} {Phys. Rev. Lett.}\ }\textbf {\bibinfo {volume} {112}},\
  \bibinfo {pages} {150802} (\bibinfo {year} {2014})}\BibitemShut {NoStop}%
\bibitem [{\citenamefont {{Nazarov}}(1999)}]{Nazarov1999}%
  \BibitemOpen
  \bibfield  {author} {\bibinfo {author} {\bibfnamefont {Y.~V.}\ \bibnamefont
  {{Nazarov}}},\ }\bibfield  {title} {\bibinfo {title} {{Universalities of weak
  localization}},\ }\href {https://doi.org/10.48550/arXiv.cond-mat/9908143}
  {\bibfield  {journal} {\bibinfo  {journal} {arXiv e-prints}\ ,\ \bibinfo
  {eid} {cond-mat/9908143}} (\bibinfo {year} {1999})}\BibitemShut {NoStop}%
\bibitem [{\citenamefont {{Nazarov}}\ and\ \citenamefont
  {{Kindermann}}(2003)}]{Nazarov_2003}%
  \BibitemOpen
  \bibfield  {author} {\bibinfo {author} {\bibfnamefont {Y.~V.}\ \bibnamefont
  {{Nazarov}}}\ and\ \bibinfo {author} {\bibfnamefont {M.}~\bibnamefont
  {{Kindermann}}},\ }\bibfield  {title} {\bibinfo {title} {{Full counting
  statistics of a general quantum mechanical variable}},\ }\href
  {https://doi.org/10.1140/epjb/e2003-00293-1} {\bibfield  {journal} {\bibinfo
  {journal} {Eur. Phys. J. B}\ }\textbf {\bibinfo {volume} {35}},\ \bibinfo
  {pages} {413} (\bibinfo {year} {2003})}\BibitemShut {NoStop}%
\bibitem [{\citenamefont {{Bachmann}}\ \emph {et~al.}(2010)\citenamefont
  {{Bachmann}}, \citenamefont {{Graf}},\ and\ \citenamefont
  {{Lesovik}}}]{Bachmann_2009}%
  \BibitemOpen
  \bibfield  {author} {\bibinfo {author} {\bibfnamefont {S.}~\bibnamefont
  {{Bachmann}}}, \bibinfo {author} {\bibfnamefont {G.~M.}\ \bibnamefont
  {{Graf}}},\ and\ \bibinfo {author} {\bibfnamefont {G.~B.}\ \bibnamefont
  {{Lesovik}}},\ }\bibfield  {title} {\bibinfo {title} {{Time Ordering and
  Counting Statistics}},\ }\href {https://doi.org/10.1007/s10955-009-9885-z}
  {\bibfield  {journal} {\bibinfo  {journal} {J. Stat. Phys.}\ }\textbf
  {\bibinfo {volume} {138}},\ \bibinfo {pages} {333} (\bibinfo {year}
  {2010})}\BibitemShut {NoStop}%
\bibitem [{\citenamefont {{Beaud}}\ \emph {et~al.}(2013)\citenamefont
  {{Beaud}}, \citenamefont {{Graf}}, \citenamefont {{Lebedev}},\ and\
  \citenamefont {{Lesovik}}}]{Lesovik13}%
  \BibitemOpen
  \bibfield  {author} {\bibinfo {author} {\bibfnamefont {V.}~\bibnamefont
  {{Beaud}}}, \bibinfo {author} {\bibfnamefont {G.~M.}\ \bibnamefont {{Graf}}},
  \bibinfo {author} {\bibfnamefont {A.~V.}\ \bibnamefont {{Lebedev}}},\ and\
  \bibinfo {author} {\bibfnamefont {G.~B.}\ \bibnamefont {{Lesovik}}},\
  }\bibfield  {title} {\bibinfo {title} {{Statistics of Charge Transport and
  Modified Time Ordering}},\ }\href {https://doi.org/10.1007/s10955-013-0815-8}
  {\bibfield  {journal} {\bibinfo  {journal} {J. Stat. Phys.}\ }\textbf
  {\bibinfo {volume} {153}},\ \bibinfo {pages} {177} (\bibinfo {year}
  {2013})}\BibitemShut {NoStop}%
\bibitem [{\citenamefont {{Tang}}\ and\ \citenamefont
  {{Wang}}(2014)}]{Tang_2014}%
  \BibitemOpen
  \bibfield  {author} {\bibinfo {author} {\bibfnamefont {G.-M.}\ \bibnamefont
  {{Tang}}}\ and\ \bibinfo {author} {\bibfnamefont {J.}~\bibnamefont
  {{Wang}}},\ }\bibfield  {title} {\bibinfo {title} {{Full-counting statistics
  of charge and spin transport in the transient regime: A nonequilibrium
  Green's function approach}},\ }\href
  {https://doi.org/10.1103/PhysRevB.90.195422} {\bibfield  {journal} {\bibinfo
  {journal} {\prb}\ }\textbf {\bibinfo {volume} {90}},\ \bibinfo {eid} {195422}
  (\bibinfo {year} {2014})}\BibitemShut {NoStop}%
\bibitem [{Note1()}]{Note1}%
  \BibitemOpen
  \bibinfo {note} {There are generalizations of DMT which preserve operators on
  $\ell $ consecutive sites \cite {Ye2020}.}\BibitemShut {Stop}%
\bibitem [{\citenamefont {{Ye}}\ \emph {et~al.}(2020)\citenamefont {{Ye}},
  \citenamefont {{Machado}}, \citenamefont {{White}}, \citenamefont {{Mong}},\
  and\ \citenamefont {{Yao}}}]{Ye2020}%
  \BibitemOpen
  \bibfield  {author} {\bibinfo {author} {\bibfnamefont {B.}~\bibnamefont
  {{Ye}}}, \bibinfo {author} {\bibfnamefont {F.}~\bibnamefont {{Machado}}},
  \bibinfo {author} {\bibfnamefont {C.~D.}\ \bibnamefont {{White}}}, \bibinfo
  {author} {\bibfnamefont {R.~S.~K.}\ \bibnamefont {{Mong}}},\ and\ \bibinfo
  {author} {\bibfnamefont {N.~Y.}\ \bibnamefont {{Yao}}},\ }\bibfield  {title}
  {\bibinfo {title} {{Emergent Hydrodynamics in Nonequilibrium Quantum
  Systems}},\ }\href {https://doi.org/10.1103/PhysRevLett.125.030601}
  {\bibfield  {journal} {\bibinfo  {journal} {\prl}\ }\textbf {\bibinfo
  {volume} {125}},\ \bibinfo {eid} {030601} (\bibinfo {year}
  {2020})}\BibitemShut {NoStop}%
\bibitem [{\citenamefont {{Cazalilla}}\ \emph {et~al.}(2011)\citenamefont
  {{Cazalilla}}, \citenamefont {{Citro}}, \citenamefont {{Giamarchi}},
  \citenamefont {{Orignac}},\ and\ \citenamefont {{Rigol}}}]{Cazalilla2011}%
  \BibitemOpen
  \bibfield  {author} {\bibinfo {author} {\bibfnamefont {M.~A.}\ \bibnamefont
  {{Cazalilla}}}, \bibinfo {author} {\bibfnamefont {R.}~\bibnamefont
  {{Citro}}}, \bibinfo {author} {\bibfnamefont {T.}~\bibnamefont
  {{Giamarchi}}}, \bibinfo {author} {\bibfnamefont {E.}~\bibnamefont
  {{Orignac}}},\ and\ \bibinfo {author} {\bibfnamefont {M.}~\bibnamefont
  {{Rigol}}},\ }\bibfield  {title} {\bibinfo {title} {{One dimensional bosons:
  From condensed matter systems to ultracold gases}},\ }\href
  {https://doi.org/10.1103/RevModPhys.83.1405} {\bibfield  {journal} {\bibinfo
  {journal} {Rev. Mod. Phys.}\ }\textbf {\bibinfo {volume} {83}},\ \bibinfo
  {pages} {1405} (\bibinfo {year} {2011})}\BibitemShut {NoStop}%
\bibitem [{\citenamefont {{Rubio-Abadal}}\ \emph {et~al.}(2019)\citenamefont
  {{Rubio-Abadal}}, \citenamefont {{Choi}}, \citenamefont {{Zeiher}},
  \citenamefont {{Hollerith}}, \citenamefont {{Rui}}, \citenamefont {{Bloch}},\
  and\ \citenamefont {{Gross}}}]{RubioAbadal2019}%
  \BibitemOpen
  \bibfield  {author} {\bibinfo {author} {\bibfnamefont {A.}~\bibnamefont
  {{Rubio-Abadal}}}, \bibinfo {author} {\bibfnamefont {J.-y.}\ \bibnamefont
  {{Choi}}}, \bibinfo {author} {\bibfnamefont {J.}~\bibnamefont {{Zeiher}}},
  \bibinfo {author} {\bibfnamefont {S.}~\bibnamefont {{Hollerith}}}, \bibinfo
  {author} {\bibfnamefont {J.}~\bibnamefont {{Rui}}}, \bibinfo {author}
  {\bibfnamefont {I.}~\bibnamefont {{Bloch}}},\ and\ \bibinfo {author}
  {\bibfnamefont {C.}~\bibnamefont {{Gross}}},\ }\bibfield  {title} {\bibinfo
  {title} {{Many-Body Delocalization in the Presence of a Quantum Bath}},\
  }\href {https://doi.org/10.1103/PhysRevX.9.041014} {\bibfield  {journal}
  {\bibinfo  {journal} {Phys. Rev. X}\ }\textbf {\bibinfo {volume} {9}},\
  \bibinfo {eid} {041014} (\bibinfo {year} {2019})}\BibitemShut {NoStop}%
\bibitem [{\citenamefont {{Kaufman}}\ \emph {et~al.}(2016)\citenamefont
  {{Kaufman}}, \citenamefont {{Tai}}, \citenamefont {{Lukin}}, \citenamefont
  {{Rispoli}}, \citenamefont {{Schittko}}, \citenamefont {{Preiss}},\ and\
  \citenamefont {{Greiner}}}]{Kaufman2016}%
  \BibitemOpen
  \bibfield  {author} {\bibinfo {author} {\bibfnamefont {A.~M.}\ \bibnamefont
  {{Kaufman}}}, \bibinfo {author} {\bibfnamefont {M.~E.}\ \bibnamefont
  {{Tai}}}, \bibinfo {author} {\bibfnamefont {A.}~\bibnamefont {{Lukin}}},
  \bibinfo {author} {\bibfnamefont {M.}~\bibnamefont {{Rispoli}}}, \bibinfo
  {author} {\bibfnamefont {R.}~\bibnamefont {{Schittko}}}, \bibinfo {author}
  {\bibfnamefont {P.~M.}\ \bibnamefont {{Preiss}}},\ and\ \bibinfo {author}
  {\bibfnamefont {M.}~\bibnamefont {{Greiner}}},\ }\bibfield  {title} {\bibinfo
  {title} {{Quantum thermalization through entanglement in an isolated
  many-body system}},\ }\href {https://doi.org/10.1126/science.aaf6725}
  {\bibfield  {journal} {\bibinfo  {journal} {Science}\ ,\ \bibinfo {pages}
  {794}} (\bibinfo {year} {2016})}\BibitemShut {NoStop}%
\bibitem [{\citenamefont {{Fukuhara}}\ \emph {et~al.}(2013)\citenamefont
  {{Fukuhara}}, \citenamefont {{Schau{\ss}}}, \citenamefont {{Endres}},
  \citenamefont {{Hild}}, \citenamefont {{Cheneau}}, \citenamefont {{Bloch}},\
  and\ \citenamefont {{Gross}}}]{Fukuhara2013}%
  \BibitemOpen
  \bibfield  {author} {\bibinfo {author} {\bibfnamefont {T.}~\bibnamefont
  {{Fukuhara}}}, \bibinfo {author} {\bibfnamefont {P.}~\bibnamefont
  {{Schau{\ss}}}}, \bibinfo {author} {\bibfnamefont {M.}~\bibnamefont
  {{Endres}}}, \bibinfo {author} {\bibfnamefont {S.}~\bibnamefont {{Hild}}},
  \bibinfo {author} {\bibfnamefont {M.}~\bibnamefont {{Cheneau}}}, \bibinfo
  {author} {\bibfnamefont {I.}~\bibnamefont {{Bloch}}},\ and\ \bibinfo {author}
  {\bibfnamefont {C.}~\bibnamefont {{Gross}}},\ }\bibfield  {title} {\bibinfo
  {title} {{Microscopic observation of magnon bound states and their
  dynamics}},\ }\href {https://doi.org/10.1038/nature12541} {\bibfield
  {journal} {\bibinfo  {journal} {Nature}\ }\textbf {\bibinfo {volume} {502}},\
  \bibinfo {pages} {76} (\bibinfo {year} {2013})}\BibitemShut {NoStop}%
\bibitem [{\citenamefont {{L{\'e}onard}}\ \emph {et~al.}(2020)\citenamefont
  {{L{\'e}onard}}, \citenamefont {{Kim}}, \citenamefont {{Rispoli}},
  \citenamefont {{Lukin}}, \citenamefont {{Schittko}}, \citenamefont {{Kwan}},
  \citenamefont {{Demler}}, \citenamefont {{Sels}},\ and\ \citenamefont
  {{Greiner}}}]{Leonard2020}%
  \BibitemOpen
  \bibfield  {author} {\bibinfo {author} {\bibfnamefont {J.}~\bibnamefont
  {{L{\'e}onard}}}, \bibinfo {author} {\bibfnamefont {S.}~\bibnamefont
  {{Kim}}}, \bibinfo {author} {\bibfnamefont {M.}~\bibnamefont {{Rispoli}}},
  \bibinfo {author} {\bibfnamefont {A.}~\bibnamefont {{Lukin}}}, \bibinfo
  {author} {\bibfnamefont {R.}~\bibnamefont {{Schittko}}}, \bibinfo {author}
  {\bibfnamefont {J.}~\bibnamefont {{Kwan}}}, \bibinfo {author} {\bibfnamefont
  {E.}~\bibnamefont {{Demler}}}, \bibinfo {author} {\bibfnamefont
  {D.}~\bibnamefont {{Sels}}},\ and\ \bibinfo {author} {\bibfnamefont
  {M.}~\bibnamefont {{Greiner}}},\ }\bibfield  {title} {\bibinfo {title}
  {{Signatures of bath-induced quantum avalanches in a many-body--localized
  system}},\ }\href {https://doi.org/10.48550/arXiv.2012.15270} {\bibfield
  {journal} {\bibinfo  {journal} {arXiv e-prints}\ ,\ \bibinfo {eid}
  {arXiv:2012.15270}} (\bibinfo {year} {2020})}\BibitemShut {NoStop}%
\bibitem [{\citenamefont {{Kinoshita}}\ \emph {et~al.}(2006)\citenamefont
  {{Kinoshita}}, \citenamefont {{Wenger}},\ and\ \citenamefont
  {{Weiss}}}]{Kinoshita2006}%
  \BibitemOpen
  \bibfield  {author} {\bibinfo {author} {\bibfnamefont {T.}~\bibnamefont
  {{Kinoshita}}}, \bibinfo {author} {\bibfnamefont {T.}~\bibnamefont
  {{Wenger}}},\ and\ \bibinfo {author} {\bibfnamefont {D.~S.}\ \bibnamefont
  {{Weiss}}},\ }\bibfield  {title} {\bibinfo {title} {{A quantum Newton's
  cradle}},\ }\href {https://doi.org/10.1038/nature04693} {\bibfield  {journal}
  {\bibinfo  {journal} {Nature}\ }\textbf {\bibinfo {volume} {440}},\ \bibinfo
  {pages} {900} (\bibinfo {year} {2006})}\BibitemShut {NoStop}%
\bibitem [{\citenamefont {{Hofferberth}}\ \emph {et~al.}(2007)\citenamefont
  {{Hofferberth}}, \citenamefont {{Lesanovsky}}, \citenamefont {{Fischer}},
  \citenamefont {{Schumm}},\ and\ \citenamefont
  {{Schmiedmayer}}}]{Hofferberth2007}%
  \BibitemOpen
  \bibfield  {author} {\bibinfo {author} {\bibfnamefont {S.}~\bibnamefont
  {{Hofferberth}}}, \bibinfo {author} {\bibfnamefont {I.}~\bibnamefont
  {{Lesanovsky}}}, \bibinfo {author} {\bibfnamefont {B.}~\bibnamefont
  {{Fischer}}}, \bibinfo {author} {\bibfnamefont {T.}~\bibnamefont
  {{Schumm}}},\ and\ \bibinfo {author} {\bibfnamefont {J.}~\bibnamefont
  {{Schmiedmayer}}},\ }\bibfield  {title} {\bibinfo {title} {{Non-equilibrium
  coherence dynamics in one-dimensional Bose gases}},\ }\href
  {https://doi.org/10.1038/nature06149} {\bibfield  {journal} {\bibinfo
  {journal} {Nature}\ }\textbf {\bibinfo {volume} {449}},\ \bibinfo {pages}
  {324} (\bibinfo {year} {2007})}\BibitemShut {NoStop}%
\bibitem [{\citenamefont {{Trotzky}}\ \emph {et~al.}(2012)\citenamefont
  {{Trotzky}}, \citenamefont {{Chen}}, \citenamefont {{Flesch}}, \citenamefont
  {{McCulloch}}, \citenamefont {{Schollw{\"o}ck}}, \citenamefont {{Eisert}},\
  and\ \citenamefont {{Bloch}}}]{Trotzky2012}%
  \BibitemOpen
  \bibfield  {author} {\bibinfo {author} {\bibfnamefont {S.}~\bibnamefont
  {{Trotzky}}}, \bibinfo {author} {\bibfnamefont {Y.~A.}\ \bibnamefont
  {{Chen}}}, \bibinfo {author} {\bibfnamefont {A.}~\bibnamefont {{Flesch}}},
  \bibinfo {author} {\bibfnamefont {I.~P.}\ \bibnamefont {{McCulloch}}},
  \bibinfo {author} {\bibfnamefont {U.}~\bibnamefont {{Schollw{\"o}ck}}},
  \bibinfo {author} {\bibfnamefont {J.}~\bibnamefont {{Eisert}}},\ and\
  \bibinfo {author} {\bibfnamefont {I.}~\bibnamefont {{Bloch}}},\ }\bibfield
  {title} {\bibinfo {title} {{Probing the relaxation towards equilibrium in an
  isolated strongly correlated one-dimensional Bose gas}},\ }\href
  {https://doi.org/10.1038/nphys2232} {\bibfield  {journal} {\bibinfo
  {journal} {Nat. Phys.}\ }\textbf {\bibinfo {volume} {8}},\ \bibinfo {pages}
  {325} (\bibinfo {year} {2012})}\BibitemShut {NoStop}%
\bibitem [{\citenamefont {{Schreiber}}\ \emph {et~al.}(2015)\citenamefont
  {{Schreiber}}, \citenamefont {{Hodgman}}, \citenamefont {{Bordia}},
  \citenamefont {{L{\"u}schen}}, \citenamefont {{Fischer}}, \citenamefont
  {{Vosk}}, \citenamefont {{Altman}}, \citenamefont {{Schneider}},\ and\
  \citenamefont {{Bloch}}}]{Schreiber2015}%
  \BibitemOpen
  \bibfield  {author} {\bibinfo {author} {\bibfnamefont {M.}~\bibnamefont
  {{Schreiber}}}, \bibinfo {author} {\bibfnamefont {S.~S.}\ \bibnamefont
  {{Hodgman}}}, \bibinfo {author} {\bibfnamefont {P.}~\bibnamefont {{Bordia}}},
  \bibinfo {author} {\bibfnamefont {H.~P.}\ \bibnamefont {{L{\"u}schen}}},
  \bibinfo {author} {\bibfnamefont {M.~H.}\ \bibnamefont {{Fischer}}}, \bibinfo
  {author} {\bibfnamefont {R.}~\bibnamefont {{Vosk}}}, \bibinfo {author}
  {\bibfnamefont {E.}~\bibnamefont {{Altman}}}, \bibinfo {author}
  {\bibfnamefont {U.}~\bibnamefont {{Schneider}}},\ and\ \bibinfo {author}
  {\bibfnamefont {I.}~\bibnamefont {{Bloch}}},\ }\bibfield  {title} {\bibinfo
  {title} {{Observation of many-body localization of interacting fermions in a
  quasirandom optical lattice}},\ }\href
  {https://doi.org/10.1126/science.aaa7432} {\bibfield  {journal} {\bibinfo
  {journal} {Science}\ }\textbf {\bibinfo {volume} {349}},\ \bibinfo {pages}
  {842} (\bibinfo {year} {2015})}\BibitemShut {NoStop}%
\bibitem [{Note2()}]{Note2}%
  \BibitemOpen
  \bibinfo {note} {If one is restricted to pure states, TEBD (as well as exact
  statevector simulations) are competitive with DMT. Unlike these, however, DMT
  can be adapted to the experimentally realistic case where the system is in a
  mixed state.}\BibitemShut {Stop}%
\bibitem [{\citenamefont {{Bertini}}\ \emph {et~al.}(2015)\citenamefont
  {{Bertini}}, \citenamefont {{De Sole}}, \citenamefont {{Gabrielli}},
  \citenamefont {{Jona-Lasinio}},\ and\ \citenamefont
  {{Landim}}}]{Bertini_2015}%
  \BibitemOpen
  \bibfield  {author} {\bibinfo {author} {\bibfnamefont {L.}~\bibnamefont
  {{Bertini}}}, \bibinfo {author} {\bibfnamefont {A.}~\bibnamefont {{De
  Sole}}}, \bibinfo {author} {\bibfnamefont {D.}~\bibnamefont {{Gabrielli}}},
  \bibinfo {author} {\bibfnamefont {G.}~\bibnamefont {{Jona-Lasinio}}},\ and\
  \bibinfo {author} {\bibfnamefont {C.}~\bibnamefont {{Landim}}},\ }\bibfield
  {title} {\bibinfo {title} {{Macroscopic fluctuation theory}},\ }\href
  {https://doi.org/10.1103/RevModPhys.87.593} {\bibfield  {journal} {\bibinfo
  {journal} {Rev. Mod. Phys.}\ }\textbf {\bibinfo {volume} {87}},\ \bibinfo
  {pages} {593} (\bibinfo {year} {2015})}\BibitemShut {NoStop}%
\bibitem [{Note3()}]{Note3}%
  \BibitemOpen
  \bibinfo {note} {We remark that either the turnstile or the related
  counting-field method is needed with TEBD in any case, to allow one to treat
  initial states that are not number-sharp.}\BibitemShut {Stop}%
\bibitem [{\citenamefont {Daley}(2014)}]{daley2014quantum}%
  \BibitemOpen
  \bibfield  {author} {\bibinfo {author} {\bibfnamefont {A.~J.}\ \bibnamefont
  {Daley}},\ }\bibfield  {title} {\bibinfo {title} {Quantum trajectories and
  open many-body quantum systems},\ }\href
  {https://doi.org/10.1080/00018732.2014.933502} {\bibfield  {journal}
  {\bibinfo  {journal} {Adv. Phys.}\ }\textbf {\bibinfo {volume} {63}},\
  \bibinfo {pages} {77} (\bibinfo {year} {2014})}\BibitemShut {NoStop}%
\bibitem [{\citenamefont {Klobas}\ \emph {et~al.}(2020)\citenamefont {Klobas},
  \citenamefont {Vanicat}, \citenamefont {Garrahan},\ and\ \citenamefont
  {Prosen}}]{klobas2020matrix}%
  \BibitemOpen
  \bibfield  {author} {\bibinfo {author} {\bibfnamefont {K.}~\bibnamefont
  {Klobas}}, \bibinfo {author} {\bibfnamefont {M.}~\bibnamefont {Vanicat}},
  \bibinfo {author} {\bibfnamefont {J.~P.}\ \bibnamefont {Garrahan}},\ and\
  \bibinfo {author} {\bibfnamefont {T.}~\bibnamefont {Prosen}},\ }\bibfield
  {title} {\bibinfo {title} {Matrix product state of multi-time correlations},\
  }\href {https://doi.org/10.1088/1751-8121/ab8c62} {\bibfield  {journal}
  {\bibinfo  {journal} {J. Phys. A: Math. Theor.}\ }\textbf {\bibinfo {volume}
  {53}},\ \bibinfo {pages} {335001} (\bibinfo {year} {2020})}\BibitemShut
  {NoStop}%
\end{thebibliography}%

\newpage


\section*{Supplementary material}

# Supplemental Material for
# "Quantum turnstiles for robust measurement of full counting statistics"

Rhine Samajdar,[1,2] Ewan McCulloch,[3] Vedika Khemani,[4] Romain Vasseur,[3] and Sarang Gopalakrishnan[5]

[1]*Department of Physics, Princeton University, Princeton, NJ 08544*
[2]*Princeton Center for Theoretical Science, Princeton University, Princeton, NJ 08544*
[3]*Department of Physics, University of Massachusetts, Amherst, MA 01003*
[4]*Department of Physics, Stanford University, Stanford, CA 94305*
[5]*Department of Electrical and Computer Engineering, Princeton University, Princeton, NJ 08544*

## SI. FULL COUNTING STATISTICS OF THE HEISENBERG SPIN CHAIN

In this section, we present additional numerical data on the FCS of the isotropic ($\Delta = 1$) Trotterized XXZ model, described by Eq. (4) of the main text. Starting from certain domain-wall initial states, in Fig. 3, we demonstrated that the mean and the variance of the transferred magnetization—in the absence of noise—point to anomalous spin transport, exhibiting a space-time scaling of $x \sim t^{2/3}$ [1–8]. In contrast, the addition of a depolarizing noise channel leads to a crossover from unitary to noisy dynamics.

Here, we extend our earlier observations to a wider range of domain-wall initial states, which are parametrized by $\mu$ as described in Eq. (5). The height of the domain wall in each instance is given by $\tanh \mu$, so, for $\mu = 0$, the initial density matrix simply describes the infinite-temperature equilibrium state. Using our turnstile-based approach, we systematically investigate the dependence of the FCS on $\mu$ using the TEBD algorithm. In our numerics, we find that the minimum bond dimension required to keep the truncation error below a fixed threshold (set to $10^{-7}$ here), progressively increases with $\mu$, the counting field, and (expectedly) the circuit depth.

Our results for varying $\mu$ are arrayed in Fig. S1: while both the mean and the variance go as $t^{2/3}$ for small $\mu > 0$, deviations from this behavior are apparent for larger $\mu \to 1$. This is in consistency with previous studies [2], which report anomalous spin diffusion for only the *high-temperature* transport. This observation of superdiffusive transport, taken together with the scaling of spin-spin correlation functions, prompted the conjecture that in the limit of $\mu \to 0$ and $t \to \infty$, the spin-transport dynamics belong to the Kardar-Parisi-Zhang universality class [3, 9, 10]. However, in order to establish the universality class of the dynamics, it is crucial to also study the higher moments, such as the skewness [Fig. S1(c)], as explored in a recent experiment with superconducting qubits [11].

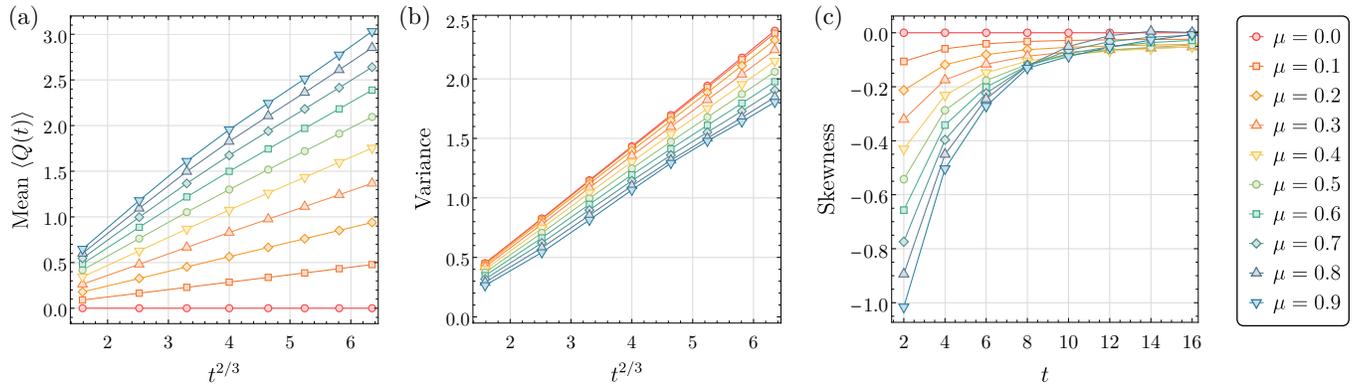

Figure S1. (a) Mean, (b) variance, and (c) skewness of the transferred magnetization for a system of 46 qubits evolving under Trotterized XXZ dynamics with $\Delta = 1$. We consider ten different domain-wall initial states parametrized by $\mu$ ranging from 0 to 0.9. As $\mu \to 0$, the mean and the variance are clearly seen to be anomalous, going as $t^{2/3}$, with obvious deviations therefrom visible at larger $\mu$.

In the main text, we also showed how the use of quantum turnstiles remains a robust approach to calculating FCS even in the presence of depolarizing noise. Here, we extend our previous analysis to address the case of amplitude damping noise as well. The noise is encoded as a quantum channel that acts on the density matrix as

$$\rho \to K_0 \rho K_0^\dagger + K_1 \rho K_1^\dagger; \quad K_0 = \begin{pmatrix} 1 & 0 \\ 0 & \sqrt{1-\gamma} \end{pmatrix}, \ K_1 = \begin{pmatrix} 0 & \sqrt{\gamma} \\ 0 & 0 \end{pmatrix} \tag{S1}$$

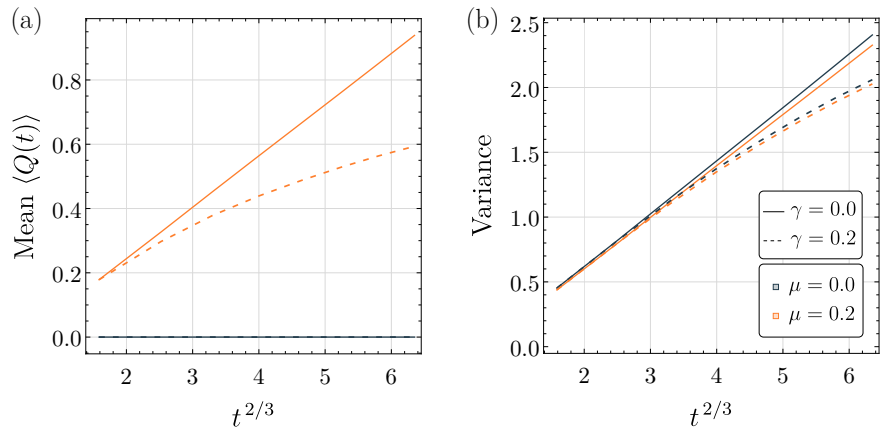

Figure S2. (a) Mean and (b) variance of $Q(t)$, for the same system as Fig. S1, in the noise-free case ($\gamma=0$, solid lines) and in the presence of amplitude-damping noise ($\gamma=0.2$, dashed lines). The colors of the individual lines encode the value of $\mu$ with gray (orange) representing $\mu=0$ ($\mu=0.2$).

where $K_0$, $K_1$ are the Kraus operators and $\gamma$ is a measure of the noise strength. In Fig. S2, we present the first two moments of the transferred magnetization as a function of time with and without amplitude damping noise. The mean and the variance both display a clear crossover from anomalous to normal transport upon turning on a nonzero $\gamma=0.2$.

## SII. CHARGE TRANSFER FULL COUNTING STATISTICS IN RANDOM CIRCUITS

In this section, we present some additional data benchmarking the turnstile approach to full counting statistics with TEBD and DMT for two far-from-equilibrium states, namely, a domain-wall state and the Néel state. We find that the "turnstile + DMT" approach enables a much less computationally expensive evaluation of the cumulant generating function of charge transfer $\chi(\lambda)$ in both nonequilibrium states.

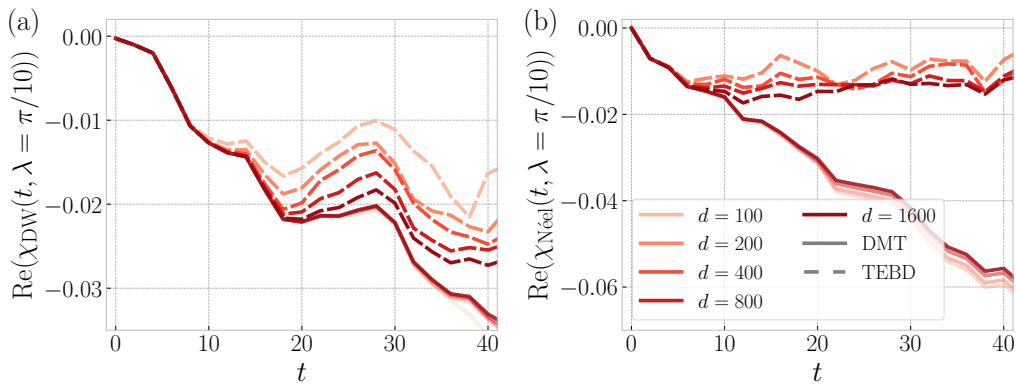

Figure S3. The cumulant generating function of charge transfer across the central bond of a random unitary circuit as a function of time, for a given value of the counting field $\lambda = \pi/10$. The two panels show $\mathrm{Re}(\chi(\lambda))$ when starting with (a) a domain-wall (DW) initial state, and (b) the Néel state.

For the domain-wall initialization, we prepare a fully polarized state $|\psi\rangle = |\uparrow \cdots \uparrow\downarrow \cdots \downarrow\rangle$ and evolve with bond dimensions $d = 100, 200, 400, 800, 1600$. With TEBD (density-matrix) evolution, the data becomes unconverged at times $t \sim 10$, whereas, with DMT, we can reach much later evolution times, $t \sim 35$. This is illustrated in Fig. S3(a) for a particular choice of the counting angle, $\lambda = \pi/10$. This improvement is comparable to that seen for the Néel state in Fig. 4 of the main text. The cumulant generating function for the Néel state, with a counting angle $\lambda = \pi/10$, is displayed in Fig. S3(b).

Lastly, we compute the charge transfer variance in different circuit realizations to confirm that the reasonably good agreement between the theoretical prediction and the "turnstile + DMT" approach is generic, and not the result of a finely tuned circuit; this is shown in Fig. S4.

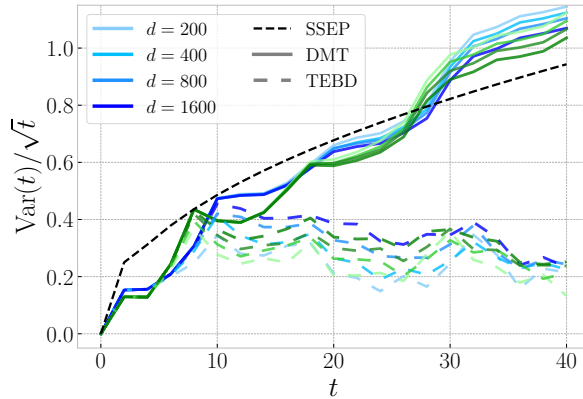

Figure S4. The variance of the charge transfer across the central bond in different random unitary circuit realizations.

[1] M. Žnidarič, Phys. Rev. Lett. **106**, 220601 (2011).
[2] M. Ljubotina, M. Žnidarič, and T. Prosen, Nat. Commun. **8**, 16117 (2017).
[3] M. Ljubotina, M. Žnidarič, and T. Prosen, Phys. Rev. Lett. **122**, 210602 (2019).
[4] S. Gopalakrishnan and R. Vasseur, Phys. Rev. Lett. **122**, 127202 (2019).
[5] J. De Nardis, M. Medenjak, C. Karrasch, and E. Ilievski, Phys. Rev. Lett. **123**, 186601 (2019).
[6] S. Gopalakrishnan, R. Vasseur, and B. Ware, Proc. Natl. Acad. Sci. U.S.A. **116**, 16250 (2019).
[7] V. B. Bulchandani, Phys. Rev. B **101**, 041411 (2020).
[8] S. Gopalakrishnan and R. Vasseur, Rep. Prog. Phys. **86**, 036502 (2023).
[9] Ž. Krajnik, E. Ilievski, and T. Prosen, SciPost Phys. **9**, 038 (2020).
[10] B. Ye, F. Machado, C. D. White, R. S. K. Mong, and N. Y. Yao, Phys. Rev. Lett. **125**, 030601 (2020).
[11] Google Quantum AI and Collaborators, "Dynamics of magnetization at infinite temperature in a Heisenberg spin chain," (2023).



\end{document}